\documentclass[twoside]{article}
\usepackage{a4wide}
\usepackage{amsmath}
\usepackage{amsfonts}
\usepackage{amssymb}
\usepackage{latexsym}
\usepackage{times}     
\usepackage{pstricks}
\usepackage{pst-grad}
\usepackage{pst-node}

\def\date{1 May 2005}

\psset{framearc=0.0,framesep=0.0truecm,nodesep=8pt,unit=0.75truecm,
      arrowsize=4pt 2,arrowinset=0.4}
\def\pstlw{0.8pt}                              
\newcounter{mycnt}
\def\themycnt{\thesection.\arabic{mycnt}}
\def\mybenv#1{\refstepcounter{mycnt}%
       \vskip 3pt\noindent{\bf #1~\themycnt}:~}
\def\myeenv{\hfill\rule{1ex}{1ex}\vskip 3pt}
\def\qed{\hfill$\Box$}
\def\ed{\end{document}}
\def\nn{\nonumber \\}
\def\Id{\text{I\!d}}
\def\antip{\textsf{S}}
\def\ip{\star}			

\def\openC{\mathbb{C}}

\def\openZ{\mathbb{Z}}
\def\grpGL{\sf{GL}}
\def\grpSp{\sf{Sp}}
\def\grpSL{\sf{SL}}
\def\grpO{\sf{O}}
\def\grpSO{\sf{SO}}
\def\grpH{\sf{H}}

\def\!{\kern -0.15ex}
\def\nn{\nonumber\\}
\def\dbrace#1#2{(\!(#1)\!)_{#2}}

\def\otp{\underline{\otimes}}
\def\ra{\rangle}
\def\la{\langle}
\def\({(\!(}
\def\){)\!)}
\def\wgt{{\rm wgt}\,}

\begin{document}
\title{\Large\sf New branching rules induced by plethysm\footnote{%
Dedicated to the memory of our co-author, friend and colleague Brian
Wybourne, 1935--2003.}}
\author{B. Fauser$^{1)}$  
\and
P.D. Jarvis$^{2)}$
\and
R.C. King$^{3)}$
\and
B.G. Wybourne$^{\dagger\,4)}$
}
{{\renewcommand{\thefootnote}{\fnsymbol{footnote}}
\footnotetext{\kern-15.3pt AMS Subject Classification:
05E05; 
16W30; 
20G10; 
11E57  
\par
\vskip 8pt\noindent\vbox{
\noindent\small\sc
1) Max Planck Institut f\"ur Mathematik, Inselstrasse 22-26, D-04103 Leipzig, 
Germany,\par\noindent\tt Bertfried.Fauser@uni-konstanz.de\par
\noindent\small\sc
2) University of Tasmania, School of Mathematics and Physics,
GPO Box 252-21, 7001 Hobart, TAS, Australia,\par 
\noindent\tt Peter.Jarvis@utas.edu.au\par
\noindent\small\sc
3) School of Mathematics, University of Southampton, 
Southampton SO17 1BJ, England,\par
\noindent\tt R.C.King@maths.soton.ac.uk\par
\noindent\small\sc
4) University of Torun, Institute of Physics, Nicholas Copernicus University,
ul.Grudziadzka 5, 87-100 Torun, Poland,\par
\noindent\tt url:~http://www.phys.uni.torun.pl/\~{}bgw \hfill\vskip 2ex 
}}}}
\maketitle
\begin{abstract}
We derive group branching laws for formal characters of subgroups 
$\grpH_\pi$ of $\grpGL(n)$ leaving invariant an  arbitrary tensor $T^\pi$
of Young symmetry type $\pi$  where $\pi$ is an integer partition. The
branchings $\grpGL(n)\downarrow\grpGL(n-1)$,  $\grpGL(n)\downarrow\grpO(n)$
and $\grpGL(2n)\downarrow\grpSp(2n)$ fixing a vector  $v_i$, a symmetric
tensor $g_{ij}=g_{ji}$ and an antisymmetric tensor
$f_{ij}=-f_{ji}$, respectively,  are obtained as special cases. All new
branchings are governed by Schur function series obtained from  plethysms
of the Schur function $s_\pi \equiv {\{} \pi {\}}$ by  the basic $M$ series of complete
symmetric functions and the $L\,=M^{-1}$ series  of elementary symmetric
functions. Our main technical tool is that of Hopf algebras, and our main
result is the derivation of a coproduct for any Schur function series
obtained by plethysm from another  such series. Therefrom one easily
obtains $\pi\! $-generalized Newell-Littlewood formulae, and the algebra of the formal
group characters of these subgroups is established. Concrete examples and extensive tabulations are
displayed for  $\grpH_{1^3}$, $\grpH_{21}$, and $\grpH_{3}$, showing their
involved and nontrivial representation theory. The nature of the subgroups
is shown to be in general affine, and in some instances non reductive.
We discuss the complexity of the coproduct formula and
give a  graphical notation to cope with it. We also discuss the way in which
the  group branching laws can be reinterpreted as twisted structures 
deformed by highly nontrivial 2-cocycles. The algebra of subgroup
characters is identified as a cliffordization of the algebra of symmetric
functions for $\grpGL(n)$ formal characters. Modification rules are beyond
the scope of the present paper, but are briefly discussed.

\end{abstract}
\eject
\tableofcontents
\section{\label{sec:Intro}Introduction}
\setcounter{equation}{0}

The use of explicit tensorial notation for handling the representation
theory of the classical groups is a natural tool in many branches of
mathematics and of mathematical or theoretical physics. The theory was
formalized by Weyl \cite{weyl:1930a}, and the associated character
techniques involving applications of symmetric functions developed
especially by Littlewood \cite{littlewood:1940a}. For modern accounts of
combinatorial representation theory, and its connections to other branches
of mathematics, we refer to the review article of Barcelo and Ram
\cite{barcelo:ram:1997a} and for the theory of symmetric functions to the
classic text of Macdonald \cite{macdonald:1979a}. A milestone in practical
applications of these techniques is the survey paper
\cite{black:king:wybourne:1983a} which gives systematic rules for handling
partition notation for labeling (finite dimensional) representations of
simple Lie groups, together with their branching rules to common subgroups,
and for the resolution of their Kronecker products. An example of
applications of Kronecker products is provided by  \cite{hasse:johnson:1993a}. A
necessary concomitant of  these techniques is the automation provided by a
symbolic  computer package such as $\textsc{Schur}^{\copyright}$
\cite{SCHUR}. Finally, some of these techniques have been found to
generalize to the representation theory of non-compact groups
\cite{king:wybourne:1985a,king:wybourne:2000a,king:wybourne:2000b}

In a recent paper \cite{fauser:jarvis:2003a} the role of symmetric
functions in relation to group representation theory has been re-considered
from the viewpoint of the underlying Hopf algebraic structure. This 
structure is in fact well known in the combinatorial literature
\cite{thibon:1991a,thibon:1991b,scharf:thibon:1994a}. In 
\cite{fauser:jarvis:2003a} the formalism of branching rules was aligned 
with certain endomorphisms on the algebra of symmetric functions, called
branching operators, derived from 1-cochains, for which the multiplicative
cohomology of Sweedler \cite{sweedler:1968a} provides a natural analytical
setting and classification. Standard branchings from generic symmetric 
functions to symmetric functions of orthogonal or symplectic type (the
classic Newell-Littlewood theorems for the group reduction from 
$\grpGL(n)\downarrow\grpO(n)$ or $\grpGL(n)\downarrow\grpSp(n)$) were found
to be derived from certain 2-cocycles (for which associativity is
guaranteed). 

This result then prompts the more general question of classifying
arbitrary, non-cohomologous, 2-cocycles, and the nature of any associated
character theory and of the algebraic or group structures which might be
entrained therewith. While an \textit{ab initio} approach to this question
is extremely difficult at this level, there is an obvious strategy for 
finding such generalizations. Namely one should look for `branching rules'
which are a direct generalization of those due to Littlewood  and for which
there is a known underlying classical  matrix group. Apart from exceptional
groups associated with specific invariants in sporadic dimensions, cases
arising from local isomorphisms, and also the
$\grpGL(n)\downarrow\grpSL(n)$ family (see section \ref{sec:Hpi} below),
such groups are necessarily not classical groups, as these are already
exhausted by the orthogonal and symplectic series. Since one studies group
characters associativity is guaranteed,  and the associated 2-cochains are
once again 2-cocycles.

This infinite reservoir of `new branching rules' is  revealed by simply
looking for formal characters associated with matrix subgroups of 
$\grpGL(n)$ which fix a certain tensor $T_\pi$ of arbitrary symmetry type, 
say corresponding to a partition ${\pi}$. The corresponding branching rules
from $\grpGL(n)$ to such subgroups thus form the topic of investigation in
the present paper.

The plan of the paper is as follows. In the section \ref{sec:2} below we
introduce the main  propositions  about coproducts of  infinite  Schur
function series,  quotients of products by such series, and products of
formal characters of subgroups, all based on plethysms of the $M\!$-series.
These results then form the basis of the calculation of `new branching
rules' from $\grpGL(n)$ to some matrix  subgroups $\grpH_\pi$ that are
exemplified in section~\ref{sec:Hpi}, with both modification rules
and product rules for $\grpH_{1^3}$ for dimension $n= 3, \,4$. Further tabulations are provided for the cases $\grpH_{21}$ and $\grpH_{3}$ in appendix~\ref{Tables}. All this is  presented  in the standard
notation  for symmetric functions,  in a purely combinatorial way. Then we
switch in section~\ref{sec:Hopf} to the much more handy and compact Hopf
algebra language, giving a Hopf algebra proof of the same results.
Thereafter we give only Hopf  proofs  in the main text, and collect the 
combinatorial proofs in appendix \ref{sec:combProofs}. Technically the new
branching rules are generated by skewing with certain infinite formal
series of symmetric functions which are generalizations of those used by
Littlewood, and some related ones needed for handling such cases as spinor
and composite tensor representations. From the analysis we also derive the $\pi\!$-generalised Newell-Littlewood product formula (Proposition \ref{Prop-pHpi}, Theorem \ref{The-munuHn}).

An underlying theme of this work is 
the use of  Littlewood's operation of plethysm 
\cite{littlewood:1958a,littlewood:1940a} of symmetric functions which 
emerges  as  pivotal  to the generation of  the new branching rules. Plethysms are defined in section
\ref{sec:2} below, and in all subsequent sections play a central role.
Specifically, in section \ref{subsec:HopfNewBranchings} key results concerning plethysms of Schur function series are derived (Main Theorems i) and ii), Theorems \ref{The:mainTheorem:i}  and \ref{The:mainTheorem ii}  respectively); in section \ref{subsec:Kernels} it is shown how plethysms allow  one  to construct an infinity of noncohomologous 2-cochains. A graphical calculus of tangle diagrams  for interpreting 
this analysis is given in appendix \ref{sec:GraphCalc}. A further study of plethysms from the viewpoint of 
the Hopf algebra structure of symmetric functions, as developed in~\cite{fauser:jarvis:2003a}, is in preparation ~\cite{fauser:jarvis:2005a}.

\noindent{\bf Dedication:} \\ 
This work is dedicated to the memory of our friend and colleague 
Brian~G.~Wybourne,  who was born on 3rd May 1935 in Morrinsville, New
Zealand and died on 26th November 2003 in Torun, Poland. It was Brian's 
typically inquisitive response to a draft copy of the `Hopf laboratory  for
symmetric functions' paper of BF and PDJ \cite{fauser:jarvis:2003a} that
encouraged him together with RCK, in the autumn of 2003 while Brian was on
one of his annual visits to Southampton, to look for examples of
non-classical Lie groups with `new' branching rules.  That discussion led
on to the present joint collaboration. We trust that our results, although
only an initial foray into a technically difficult area of combinatorial
representation theory, are in keeping with Brian's vision for some of the
last research that occupied him before his untimely death.

\section{\label{sec:2}Schur functions and Schur function series} 
\setcounter{equation}{0}

\subsection{Basic notions}

As mentioned in the introduction, in this work we are dealing with formal
group characters, which generically may be regarded as particular sorts of
symmetric functions. Abstractly these   lie in the ring of invariant 
polynomials $\Lambda^n={\openZ}[x_{1},x_{2},\ldots,x_{n}]^{S_{n}}$ in
indeterminates $x_{1}, x_{2}, \ldots, x_{n}$ which are symmetric in  their
arguments. Via an inductive limit one passes over to infinitely many
variables obtaining a graded ring $\Lambda=\oplus \Lambda^n$. As is well
known, there are many bases for symmetric functions whose products and sums
suffice to build up arbitrary symmetric polynomials 
\cite{macdonald:1979a}. Of particular interest for this paper are the
so-called Schur functions. With an appropriate interpretation of
$x=(x_1,x_2,\ldots,x_n)$, the Schur function $s_\lambda(x)$, or 
$\{\lambda\}$ in the notation of Littlewood \cite{littlewood:1940a}, is 
the character of the irreducible representation $V^\lambda$ of  $\grpGL(n)$
of highest weight $\lambda$. In this notation $\lambda$ is an integer
partition, that is $\lambda =(\lambda_{1}, \lambda_{2}, \ldots,
\lambda_{\ell})$, an ordered $\ell$-tuple of positive integers such that
$\lambda_{1} \ge\lambda_{2} \ge \ldots \ge \lambda_{\ell}$; $\lambda$ is a
partition of $\omega_{\lambda} \equiv |\lambda| := \sum_{i} \lambda_{i}$,
and the number of parts or \textit{length} of the partition is
$\ell(\lambda)$ or just $\ell$ in this case.

The ring of symmetric functions $\Lambda ={\openZ}[x_{1},x_{2},\ldots]$ 
admits various operations, such as addition and  (outer) multiplication. 
The latter gives rise in the Schur function basis to the  famous
Littlewood-Richardson  coefficients as the structure constants in the
product
\begin{align}
s_\lambda(x) \cdot s_\mu(x) 
&=\sum_{\nu} {C_{\lambda\mu}}^{\nu} s_\nu(x).
\end{align}
Reciprocally, there is an adjoint (outer) skew operation such that,
\begin{align}
s_{\nu/\mu}(x)
&=\sum_{\lambda} {C_{\lambda\mu}}^{\nu} s_\lambda(x).
\end{align}
Additional symmetric function definitions will be introduced in the sequel
as they arise. It should be noticed that we are flexible as to whether the
discussion covers \textit{universal characters}, where the number of
indeterminates is formally infinite. Then the polynomial ring is being
extended to the closure using the inductive limit of letting the number of
variable tend to infinity. Doing that, one knows that so-called syzygies are avoided, and
such formal representation modules are flat. By contrast for symmetric
functions with a finite number $n$ of variables,  for example model subgroups of
$\grpGL(n)$, one hits the problem of dealing with such syzygies. This
amounts to introducing further relations, the so called modification
rules. A systematic treatment of
modification rules for non classical groups is beyond the scope of
the paper, but we will exhibit a few instances to
show how to cope with them in  practice. Generally speaking our aim is to
display a variety of examples for specific matrix groups, for which a
finite number of  variables is necessary and intended. However, one should
keep in mind that abstract arguments, for example in section
\ref{sec:Hopf}, are dealt with in the infinite variable case. This is
common practice in the classical theory too.

\subsection{\label{sec:plethysm}Plethysm}

One more advanced piece of symmetric function formalism, that of plethysm,
is central to the structure of Schur function series and branching rules
old and new. We give notation and some formal definitions here. We denote
Schur functions as $s_\lambda$, or $\{\lambda\}$ in Littlewood's bracket
notation. 

\subsubsection{\label{sec:PlethComp}Plethysm as composition}

The mathematical definition of plethysm is given by the composition of
Schur functions, $s_\mu[s_\lambda]$. Using Littlewood's notation for Schur
functions $s_\lambda\equiv \{\lambda\}$, it is  customary to write a
plethysm using the symbol for a tensor product $s_\mu[s_\lambda] \equiv
\{\lambda\}\otp\{\mu\}$ with reversed terms.  We use the underlined tensor
product symbol $\otp$ for plethysm, to make a distinction  with the tensor
sign $\otimes$ appearing in the Hopf algebra development below. We do not
use the  $\circ$ as symbol for composition either, since this symbol is
frequently employed for inner products, which we however denote by $\ip$.
The `composition' definition of plethysm may be explained as follows  in
the context of the formal combinatorial definition of a Schur function
\cite{macdonald:1979a}. A Schur function is given by 
\begin{align}
s_\lambda(x) 
&=\sum_{T \in ST^{\lambda}} x^{\wgt(T)},
\end{align}
where the sum is over all tableaux (fillings) $T$ belonging to the set
$ST^{\lambda}$ of semi-standard tableaux of shape $\lambda$. Each summand
is a monomial in the variables $x_{1}, x_{2}, \ldots, x_{n}$. If there are
$m$ such monomials in the Schur function $s_\lambda(x)$, and these are
denoted by $y_{i}$ with $i=1,2,\ldots,m$, then the plethysm, or 
composition, of the Schur function $s_\mu$ with the Schur function
$s_\lambda$, is given by
\begin{align}
s_\mu[s_\lambda](x)=s_\mu(y)
&=\sum_{T \in ST^{\mu}} y^{\wgt{(T)}}
\end{align}
where the entries in each tableau are now  taken from the set
$\{y_{i}\,|\,i=1,2,\ldots,m\}$ of monomials of the Schur function
$s_\lambda(x)$.
\mybenv{Example} 
Consider $s_{(2)}[s_{(1^2)}](x_1,\ldots,x_4)$. Expand $s_{(1^2)}$ as
\begin{align}
s_{(1^2)}(x_1,\ldots,x_4)
&=x_{1}x_{2}+x_{1}x_{3}+x_{1}x_{4}+x_{2}x_{3}+x_{2}x_{4}+x_{3}x_{4} \nn
&=y_{1}+y_{2}+y_{3}+y_{4}+y_{5}+y_{6}
\end{align}
which leads to the expansion of the composition $s_{(2)}[s_{(1^2)}]$ as
\begin{align}
s_{(2)}[s_{(1^2)}](x_{1},\ldots,x_{4})
&=s_{(2)}(y_{1},\ldots,y_{6}) 
 =s_{(2)}(x_{1}x_{2},\ldots,x_{3}x_{4}) \nn
&= y_1^2 + \cdots + y_6^2 + y_1y_2 + \cdots +  y_5y_6 \nn  
&= x_1^2x_2^2 + \cdots + x_3^3x_4^2 + x_1^2x_2x_3 +\cdots + x_4^2x_2x_3 
    + 3x_1x_2x_3x_4 \nn
&=
 s_{(2^2)}(x_{1},\ldots,x_{4})
+s_{(1^4)}(x_{1},\ldots,x_{4})
\end{align}
Here the problem of the evaluation of the plethysm is  simply to expand
$s_{(2)}(y)$  in the Schur function basis  $s_\nu(x)$ with $\nu$ a
partition of $4$.
\myeenv

\subsubsection{Plethysms related to branchings}

Consider two groups $\grpGL(m)$, $\grpGL(n)$, with $m>n$. Consider a Schur
function $\{\lambda\}$ which represents the  character of an irreducible
$m$-dimensional representation of $\grpGL(n)$. This representation can be
surjectively embedded in the fundamental representation of the group
$\grpGL(m)$ whose character is $\{1\}$. The branching process
$\grpGL(m)\rightarrow\grpGL(n)$ is then described by the injective map
$\{1\}\rightarrow\{\lambda\}$ which leads to the general formula
\begin{align}
\grpGL(m)\rightarrow\grpGL(n)\,:
&&
\{\mu\}\rightarrow \{\lambda\} \otp \{\mu\}
\end{align} 

This process, intimately related to physics, is the origin of the usage of
the tensor symbol $\otp$.  The connection with the previous definition of a
plethysm as a composition comes about because the $\grpGL(n)$ character 
$\{\lambda\}$ is nothing other than the Schur function $s_\lambda(x)$, with
a suitable identification of $x$. Its dimension $m$, obtained  by
setting all $x_i=1$, is just the number of monomials $y_i$ in
$s_\lambda(x)$, and $s_\mu(y)$ is the corresponding $\grpGL(m)$ character
$\{\mu\}$.

In terms of modules, let $V^\lambda$ be the $\grpGL(n)$-module with
character $\{\lambda\}$ and dimension $m$. This module may be identified
with the defining $\grpGL(m)$-module $V$ on which $\grpGL(m)$ acts
naturally.  Then the plethysm $\{\lambda\}\otp\{\mu\}$ arises as the
character of the  $\grpGL(m)$-module $V^\mu=(V^\lambda)^\mu$ viewed as a
$\grpGL(n)$-module. As a result of this interpretation it is sometimes
convenient to adopt a notation for plethsyms whereby the corresponding
character is denoted  not by  $\{\lambda\} \otp \{\mu\}$, but by
$\{\lambda\}^{\otp\{\mu\}}$.

\subsubsection{Plethysms and outer exponentiation}

Plethysm can be tied to the following problem of invariants of  matrices
$\vert A^{\lambda}\vert$. For example the characteristic polynomial gives a
relationship between the roots $x_i$, that is  invariants, and the
coefficients of the polynomials. The process of evaluating a plethysm is
the same as that of computing  the coefficients of the characteristic
polynomial having roots $x_i^k$ from the coefficients of the original
polynomial. In general one tries to compute the  invariant $\vert
A^{\lambda}\vert^{\mu}$  where $\mu$ is a partition of $k$. The relation to
outer product tensor powers is then as follows. The plethysm
\begin{align}
\{\lambda\}\otp\{\mu\} &= \sum_\nu p_{\lambda,\mu}^{\nu}\{\nu\},
\end{align} 
with non-negative integers $p_{\lambda,\mu}^{\nu}$, appears  in the outer
tensor product $k$-fold power of the Schur function $\{\lambda\}$ given by
\begin{align}
\{\lambda\}^k &=\sum_{\mu,|\mu|=k}
f^\mu\ \{\lambda\}\otp\{\mu\},
\end{align}
where the multiplicity $f^\mu$ is  the dimension of the irreducible
representation $\{\mu\}$ of the symmetric group $S_k$. This makes it clear
that the plethysm $\{\lambda\}\otp\{\mu\}$ is nothing other than the
character of the $\{\mu\}$-symmetrised  tensor power of the
$\grpGL(n)$-module $V^\lambda$, justifying yet  again the exponential
notation $\{\lambda\}^{\otp\{\mu\}}$ that we will meet in our main
theorem.  As iterated outer multiplication,  that is exponentiation, the
operation of plethysm is of course not commutative, and satisfies various
forms of right and left  distributivity, which in fact can be used to
compute plethysms  iteratively. Details are deferred to section
\ref{subsec:HopfNewBranchings} below, where  the results  are needed.

\subsection{\label{sec:SSeries}Schur function series}

Littlewood~\cite{littlewood:1940a} introduced a set of infinite Schur
function series which, much later allowed  King~\cite{king:1975} to
formulate various identities  and branching rules  in an extremely compact
notation.  These identities have been extended by King, Dehuai and Wybourne
\cite{king:dehuai:wybourne:1981a} and later by Yang and Wybourne
\cite{yang:wybourne:1986a}; we follow the presentation of the
latter\footnote{%
There are misprints in eqns. (2), (3) of \cite{yang:wybourne:1986a}
and a weak notation which unfortunately was kept in
\cite{fauser:jarvis:2003a}. To cope notationally with the transformations
$\tilde{~}$, $\dagger$, ${~}^{-1}$, one needs to consider the formal 
parameter $t$ of the series.
}.

A Schur function series is an infinite series of Schur functions often
given via a generating function defining a formal power series. The most
basic  Schur function series are the mutually inverse series $L$ and 
$M=L^{-1}$.
\begin{align}
L_t(x) = \prod_{i=1}^\infty (1-x_i\,t) 
&& 
M_t(x) = \prod_{i=1}^\infty (1-x_i\,t)^{-1} = L_t^{-1}(x)
\end{align}
from which the others may be derived using plethysm, sum and product. The
Schur function content of these mutually inverse series is
\begin{align}
L_t(x) = \sum_{m=0}^\infty (-1)^{m} s_{(1^m)}(x)\,t^m =
         \sum_{m=0}^\infty (-1)^{m} \{ 1^m \}\,t^m, 
&& 
M_t(x) = \sum_{m=0}^\infty  s_{(m)}(x)\,t^m =
         \sum_{m=0}^\infty  \{ m \}\,t^m \, ,
\end{align}
again using the notation $\{\lambda\}$ of Littlewood for a Schur function
$s_\lambda(x)$. 

Furthermore it is convenient to follow Yang and Wybourne to introduce the
conjugate (with respect to transposed partitions) series,  signified by
$\tilde\,$, and the inverse conjugate or adjoint series, signified by
$\dagger$:
\begin{align}
P_t(x)&=\tilde{L}_t(x)=L_{-t}(x)^{-1}
 \,=\, \prod_{i=1}^\infty (1+x_i\,t)^{-1}
 \,=\, \sum_{m=0}^\infty (-1)^m\,\{m\}\,t^m\,,\nn
Q_t(x)&=L_t^\dagger(x)=(\tilde{L}_t(x))^{-1}=\widetilde{L_t^{-1}}(x)=
L_{-t}(x)
\,=\, \prod_{i=1}^\infty (1+x_i\,t) 
 \,=\, \sum_{m=0}^\infty \{1^m\}\,t^m\,.
\end{align}
Notice that taking the adjoint is equivalent to the transformation $t
\rightarrow -t$, while the inversion $L \rightarrow L^{-1}$ can be viewed
as a plethysm.
\begin{align}
L_t^{-1}(x) &= L_t(-x) \,=\, (-\{1\})\otp L_t \nn
\tilde{L}_t(x) &= L_{-t}(-x) \,=\, (-\{1\})\otp L_{-t}.
\end{align} 
The other series\footnote{%
The series can be considered to belong to the ring of formal power series 
$\Lambda[[t]]$ associated to the ring of symmetric functions $\Lambda$,
which admits a $\lambda$-ring structure 
\cite{macdonald:1979a,knutson:1973a}. 
} 
and their forms as plethysms are then derived in a similar manner, see
\cite{yang:wybourne:1986a}. We may be lax about the index $t$ since we are
dealing mainly with the  series $L$ and  $M$. Generally,  Schur function
series come in pairs which are mutually inverse and consecutively named.
One finds
\begin{align}
AB=CD=EF=GH=LM=PQ=RS=VW=1 \,.
\end{align}
Generaing functions and relations between some of these are  displayed in
table 2, following \cite{yang:wybourne:1986a}. While we are following in
this section the habit found in the literature to derive Schur function series
from the $L$ series, the forthcoming sections are based on the $M$ series
to avoid frequent usage of the clumsy notation $L^{-1}$. 

\begin{table}[ht]\begin{center}
\caption{Schur function series: Type, Name=Product, Schur function content, and
plethysm.}
\begin{align}
\label{eq:series}
\begin{array}{c@{~~}c@{\,=\,}l@{~~}l@{~~}l@{~~}l}
\hline\hline\\
L      & L_t & \prod_i(1-x_it) & \sum_m(-1)^m\{1^m\}t^m & L_t(x_i)
& \{1\}\otp L_t \\
L^{-1} & M_t & \prod_i(1-x_it)^{-1}  & \sum_m\{m\}t^m & L_t(x_i)^{-1}
& (-\{1\})\otp L_t\\
L^\dagger & Q_t & \prod_i(1+x_it) & \sum_m \{1^m\}t^m & L_{-t}(x_i)
& \{1\}\otp L_{-t}\\
\tilde{L} & P_t & \prod_i(1+x_it)^{-1}  & \sum_m (-1)^m\{m\}t^m & 
L_{-t}(x_i)^{-1} & (-\{1\})\otp L_{-t}\\
A & A_t & \prod_{i<j}(1-x_ix_jt) & \sum_\alpha (-1)^{\omega_\alpha/2}
\{\alpha\}t^{\omega_\alpha} & L_t(x_ix_j) (i<j)
& \{1^2\}\otp L_t\\
A^{-1} & B_t & \prod_{i<j}(1-x_ix_jt)^{-1} & \sum_\beta
\{\beta\}t^{\omega_\beta} &
L_t(x_ix_j)^{-1} (i<j)
& (-\{1^2\})\otp L_t\\
\tilde{A} & C_t & \prod_{i\le j}(1-x_ix_jt) & \sum_\gamma
(-1)^{\omega_\gamma/2}\{\gamma\}t^{\omega_\gamma} & L_t(x_ix_j) (i \le j)
& \{2\}\otp L_t\\
A^\dagger & D_t & \prod_{i\le j}(1-x_ix_jt)^{-1} &
\sum_\delta \{\delta\}t^{\omega_\delta} & L_t(x_ix_j)^{-1} (i \le j)
& (-\{2\})\otp L_t\\
V=\tilde{V} & V_t & \prod_{i}(1-x_i^2t) &
\sum_{p,q} (-1)^p\{\widetilde{p+2q},p\}t^{p+q} & L_t(x_i^2)
& (\{2\}-\{1^2\}) \otp L_t\\
V^{-1}=V^\dagger & W_t & \prod_{i}(1-x_i^2t)^{-1} &
\sum_{p,q} (-1)^p \{ p+2q,p \}t^{p+q} & L_t(x_i^2)^{-1}
& (\{1^2\}-\{2\})\otp L_t\\[1ex]
\hline
\end{array}
\end{align}
\end{center}\end{table} 

Some remaining series are $E=LA$, $F=L^{-1}A^{-1}$, $G=L^\dagger A$,
$H=\tilde{L}A^{-1}$, $R=L\tilde{L}$, $S=L^{-1}L^\dagger$,
$V=\tilde{A}A^{-1}$ and $W=AA^\dagger$.  In the table the following are the
characterising properties of the partitions that have been used: 
$(\alpha)$ is given in Frobenius notation by,
\begin{align}
(\alpha) &= \left(\begin{array}{cccc}
a_1 & a_2 & \ldots & a_r \\
a_1+1 & a_2+1 & \ldots & a_r+1
\end{array}\right)
\end{align}
and $\{\gamma\}$ is its conjugate (obtained in Frobenius notation  by
interchanging the two rows); $\{\delta\}$ has only even parts and
$\{\beta\}$ is its conjugate; $\{\epsilon\}$ (not in the table, but related
to $E$) is self conjugate; $\{\zeta\}$  (not in the table, but related to
$F$) is an arbitrary partition. 

We will use $\(\ldots\)_\pi$ as character brackets for the infinitely  many
new types of characters derived using $M_\pi=\{\pi\}\otp M$ series in the
sequel. Often, the index $\pi$ is replaced by the  dimension of the
representation $\(\ldots\)_{\textrm{dim}}$, but confusion should not occur.

The branching rules for the restriction from $\grpGL(n)$ to the subgroups 
$\grpGL(n-1)$, $\grpO(n)$ and $\grpSp(n)$ (for $n$ even) are given by
skewing the Schur functions corresponding to characters of $\grpGL(n)$ by
various infinite Schur function series,
\begin{align}
\grpGL(n)\supset \grpGL(n-1)&&
	\{\lambda\}\rightarrow\{\lambda/(\{1\}\otp M)\}
	&=\{\lambda/M\}=\(\lambda/M_{1}\)_{\{1\}},
	\label{Eq-brGLn-1}\\
\grpGL(n)\supset \grpO(n)&&  
	\{\lambda\}\rightarrow[\lambda/(\{2\}\otp M)]
	&=[\lambda/D]=\(\lambda/M_{2}\)_{\{2\}},
	\label{Eq-brOn}\\
\grpGL(n)\supset \grpSp(n)&& 
	\{\lambda\}\rightarrow\la\lambda/(\{1^2\}\otp M)\ra
	&=\la\lambda/B\ra=\(\lambda/M_{1^2}\)_{\{1^2\}}.
	\label{Eq-brSpn}
\end{align} 

The origin of these rules is the fact that $\grpGL(n-1)$, $\grpO(n)$ and 
$\grpSp(n)$ are the subgroups of $\grpGL(n)$ that leave invariant a vector,
a symmetric second rank tensor, and an antisymmetric second rank tensor, 
or in a basis for ${\openC}^{n}$, objects $v_i$, $g_{ij} = g_{ji}$,
$f_{ij} = -f_{ji}$, respectively. It is natural to ask what are the
subgroups of $\grpGL(n)$ that leave invariant higher rank tensors, such
as a third rank fully antisymmetric tensor\footnote{%
There seem to be only a few instances of research on this or related 
topics, for example \cite{dubois-violette:henneaux:2002a}.}
$\eta_{ijk} = - \eta_{jik} = -\eta_{ikj}$ would have Young type $\{1^3\}$.
If we denote the corresponding subgroup\footnote{%
In principle can consider subgroups which leave invariant a linear
combination of tensors of different Young symmetry type, \emph{e.g.} 
${\{}1^{2}{\}}{\{}1{\}} = {\{}1^{3}{\}}+ {\{}21{\}}$, but we postpone to
investigate this complication until section \ref{sec:Hopf}. }
by $\grpH_{1^3}(n)$, then the corresponding branching rule is formally
given by
\begin{align}
\label{Eq-brHn}
\grpGL(n)\supset \grpH_{1^3}(n)&&
\{\lambda\}\rightarrow\(\lambda/(\{1^3\}\otp M)\)
&=\(\lambda/M_{1^3}\)_{1^3}.
\end{align}
More generally, if $\grpH_\pi(n)$ is the subgroup of $\grpGL(n)$ leaving
invariant a tensor whose symmetry is specified by the partition  $\pi$, 
we have\footnote{%
Analogously to branching by skewing, formal multiplicative branchings, for
example $\{\lambda\} \mapsto \( \lambda \cdot M \)$, $\{\lambda\}\mapsto \(
\lambda \cdot D \)$, $\{\lambda\}\mapsto \( \lambda \cdot B \)$, arise in
the case of the branching of unitary representations of non-compact Lie
groups \cite{yang:wybourne:1986a,king:wybourne:2000a,king:wybourne:2000b}
to representations of a maximal compact Lie subgroup.  An analogous
generalisation of multiplicative branching, for example $\{\lambda\}\mapsto
\( \lambda \cdot \{ \pi \} \otp M \)$ might equally be considered.}
\begin{align}
\label{Eq-brHpi}
\grpGL(n)\supset \grpH_\pi(n)&&
\{\lambda\}\rightarrow\(\lambda/(\{\pi\}\otp M)\)_\pi
&=\(\lambda/M_\pi\)_\pi.
\end{align}
These formal identities may be inverted through the use of $L=M^{-1}$ to
give
\begin{align}
\grpGL(n-1)\subset \grpGL(n)&&  
	\{\lambda\}\rightarrow\{\lambda/(\{1\}\otp L)\}
	&=\{\lambda/L\}=\{\lambda/M_1^{-1}\},
	\label{Eq-brinvGLn-1}\\
\grpO(n)\subset \grpGL(n)&&  
	[\lambda]\rightarrow\{\lambda/(\{2\}\otp L)\}
	&=\{\lambda/C\}=\{\lambda/M_{2}^{-1}\},
	\label{Eq-brinvOn}\\
\grpSp(n)\subset \grpGL(n)&&
	\la\lambda\ra\rightarrow\{\lambda/(\{1^2\}\otp L)\}
	&=\{\lambda/A\}=\{\lambda/M_{1^2}^{-1}\},
	\label{Eq-brinvSpn}\\
\grpH_{1^3}(n)\subset \grpGL(n)&&
	\(\lambda\)\rightarrow\{\lambda/(\{1^3\}\otp L)\}
	&=\{\lambda/M_{1^3}^{-1}\},\,
	\label{eq-brinvHn}
\end{align}
and, more generally, for the subgroup of type $\grpH_\pi(n)$,
\begin{align}
\grpH_\pi(n)\subset \grpGL(n)&&
	\(\lambda\)_\pi \rightarrow\{\lambda/(\{\pi\}\otp L)\}
	&=\{\lambda/L_\pi\}=\{\lambda/M_\pi^{-1}\}.
\label{Eq-brinvHpi}
\end{align}
In the following subsections we give some of the systematics of general 
series $\Phi$, including $M_\pi$ and their inverses $M_\pi^{-1}$ as special
cases. These series encode the complexity necessary to deal with  the
manipulation of formal characters $\{\lambda\}_\Phi$, including as a
special case for the $M_\pi$ series, the formal characters
$\(\lambda\)_\pi$. Having done this, we will be in a  position to examine
specific cases of non-classical subgroups $H_\pi$ as an  illustration of
the general theme of the paper. In the course of this work, we pause to
recall a few notions about the outer Hopf algebra structure of  symmetric
functions. Switching to a more advanced description will ease the
presentation and proofs; appendix \ref{sec:combProofs} provides the 
combinatorial details for the more conservative reader.

\subsection{\label{sec:Heuristics}``Coproducts'' of Schur function
series: combinatorics}

As will be seen below, a crucial part of the manipulations with  branching
rules is associated with re-writing a symmetric function of  a set of
indeterminates $z = (z_1,z_2,\ldots)$  in terms of symmetric functions of
its parts, if it is regarded as  partitioned into two subsets $x =
(x_1,x_2,\ldots)$ and  $y = (y_1,y_2,\ldots)$, or in the finite-dimensional
case  $z=(x,y)=(x_1,x_2,\ldots,x_m,y_1,y_2,\ldots,y_n)$. For reasons to be
explained below, this expansion is here called a `coproduct'. For the
moment we simply proceed with the  explicit steps.

\mybenv{Proposition} 
It is well known that
\begin{align}
  M(x,y)&=M(x)M(y),
\label{Eq-coM} \\
  L(x,y)&=L(x)L(y),
\label{Eq-coL} \\   
  A(x,y)&=A(x)A(y)\sum_{\sigma}(-1)^{|\sigma|}\ s_\sigma(x)\ s_{\sigma'}(y),
\label{Eq-coA} \\
  B(x,y)&=B(x)B(y)\sum_{\sigma}\ s_\sigma(x)\ s_\sigma(y),
\label{Eq-coB} \\
  C(x,y)&=C(x)C(y)\sum_{\sigma}(-1)^{|\sigma|}\ s_\sigma(x)\ s_{\sigma'}(y),
\label{Eq-coC} \\
  D(x,y)&=D(x)D(y)\sum_{\sigma}\ s_\sigma(x)\ s_\sigma(y),
\label{Eq-coD} 
\end{align}
where $|\sigma|\equiv \omega_\sigma$ is the weight of the partition 
$\sigma$, and $\sigma'$ denotes the conjugate of $\sigma$.
\myeenv

\noindent{\bf Proof:}
The first of these results (\ref{Eq-coM}) corresponds to the trivial
observation that
\begin{align}
  M(x,y)&=\prod_{i=1}^m (1-x_i)^{-1} \prod_{a=1}^n
  (1-y_a)^{-1}=M(x)M(y).
\label{Eq-Mxy}
\end{align}
A similar observation immediately gives (\ref{Eq-coL}). The derivations of
(\ref{Eq-coA})-(\ref{Eq-coD}) depend on the Cauchy identity and its inverse:
\begin{align}
{\sf C}(x,y)
\,=\,\prod_{{1\leq i\leq m}\atop{1\leq a\leq n}}\ 
(1-x_iy_a)^{-1}
 &=\sum_{\sigma}\  s_\sigma(x)\ s_\sigma(y),
\label{Eq-Cauchy}\\
{\sf C}^{-1}(x,y)
\,=\,\prod_{{1\leq i\leq m}\atop{1\leq a\leq n}}\ 
(1-x_iy_a)
  &=\sum_{\sigma}\  (-1)^{|\sigma|}\ s_\sigma(x)\ s_{\sigma'}(y).
\label{Eq-invCauchy}
\end{align}
These can be used to give, for example,
\begin{align}
B(x,y)
&=\prod_{1\leq i<j\leq m} (1-x_ix_j)^{-1} 
\prod_{{1\leq i\leq m}\atop{1\leq a\leq n}} (1-x_iy_a)^{-1}
\prod_{1\leq a<b\leq n} (1-y_ay_b)^{-1} \nn
&=
B(x)\ \left(\sum_{\sigma} s_\sigma(x)\ s_\sigma(y)\right)\  B(y),
\end{align}
thereby proving (\ref{Eq-coB}). The remaining results follow in the same
way.\qed

\noindent
Similarly, using (\ref{Eq-Cauchy}) more than once in the case
of $M_{1^3}(x)$ we find
\begin{align}
M_{1^3}(x,y)
&=
M_{1^3}(x)\,M_{1^3}(y) \sum_{\sigma,\tau} 
  s_{\sigma}(x)\ s_{\{1^2\}\otp\{\tau\}}(x)
  \ s_{\{1^2\}\otp\{\sigma\}}(y)\ s_\tau(y).
\label{Eq-coU}
\end{align}
More generally we can formulate the important
\mybenv{Proposition}\label{Prop-Mpixy}
For any partition $\pi$
\begin{align}
M_\pi(x,y)
&=M_\pi(x)\,M_\pi(y)\ 
  \prod_{\xi,\eta<\pi}\ 
  \prod_{k=1}^{C^\pi_{\xi\eta}}\
  \sum_{\sigma(\xi,\eta,k)}\
  s_{\xi\otp\sigma(\xi,\eta,k)}(x)\ s_{\eta\otp\sigma(\xi,\eta,k)}(y),
\label{Eq-Mpixy}
\end{align}
where the coefficients $C^\pi_{\xi\eta}$ are the Littlewood-Richardson
coefficients defined by
\begin{align}
s_\xi(x)\ s_\eta(x)&= \sum_\pi\ C^\pi_{\xi\eta}\ s_\pi(x),
\label{Eq-outerLR}
\end{align}
or, equivalently,
\begin{align}
s_\pi(x,y) 
&=
\sum_{\xi,\eta}\ C^\pi_{\xi\eta} \ s_\xi(x)\ s_\eta(y).
\label{Eq-coLR}
\end{align} 
\myeenv
\noindent{\bf Proof:} 
Let $N^\pi=\sum_{\xi\eta}\ C^\pi_{\xi\eta}$ be the number of summands 
in the expansion of $s_\pi(x,y)$ in the form (\ref{Eq-coLR}). This 
includes the two summands $s_\pi(x)$ and $s_\pi(y)$, which may conveniently
be taken to be the first and the last, respectively, corresponding to the
cases $(\xi,\eta)=(\pi,0)$ and $(\xi,\eta)=(0,\pi)$, which occur with
multiplicity $1$. Then
\begin{align}
s_\pi(x,y)
&=\sum_{\xi,\eta}\ C^\pi_{\xi\eta} \ s_\xi(x)\ s_\eta(y)
 =\sum_{k=1}^{N^\pi}\ s_{\xi(k)}(x)\ s_{\eta(k)}(y).
\label{Eq-spixy}
\end{align}
It follows that
\begin{align}
M_\pi(x,y)
&=\prod_{T\in  ST^\pi}\  
       \left(1-(x,y)^{\wgt(T)}\right)^{-1}\nn
&=\prod_{k=1}^{N^\pi}\ 
       \prod_{U\in ST^{\xi(k)}}\ 
       \prod_{V\in ST^{\eta(k)}}\
       \left(1-x^{\wgt(U)}\,y^{\wgt(V)}\right)^{-1}\nn  
&= \prod_{k=1}^{N^\pi} \sum_{\sigma(k)}\
       s_{\sigma(k)}[s_{\xi(k)}](x)\ s_{\sigma(k)}[s_{\eta(k)}](y)\nn
&= M_\pi(x)\ M_\pi(y)\ 
     \prod_{k=2}^{N^\pi-1}\ 
     \sum_{\sigma(k)}\
     s_{\xi(k)\otp\sigma(k)}(x)\ s_{\eta(k)\otp\sigma(k)}(y)\nn
&= M_\pi(x)\,M_\pi(y)\ 
\prod_{\xi,\eta<\pi}\ 
  \prod_{k=1}^{C^\pi_{\xi\eta}}\
  \sum_{\sigma(\xi,\eta,k)}\
  s_{\xi\otp\sigma(\xi,\eta,k)}(x)\ s_{\eta\otp\sigma(\xi,\eta,k)}(y),
\end{align}
where in the second step use has been made of Cauchy's identity
(\ref{Eq-Cauchy}), while in the third it has been assumed that
$k=1$ and $k=N^\pi$ correspond to the two summands $s_\pi(x)$ 
and $s_\pi(y)$, respectively.\qed

\mybenv{Example}\label{Ex:DMpi}
It follows from Proposition~\ref{Prop-Mpixy} that:
\begin{align}
M_3(x,y)
&=M_3(x)\,M_3(y)\ \sum_{\sigma,\tau} 
  s_{\sigma}(x)\ s_{\{2\}\otp\tau}(x)\cdot s_{\{2\}\otp\sigma}(y)\ s_\tau(y),
\label{Eq-M3xy} \\
M_{21}(x,y)
&=M_{21}(x)\,M_{21}(y)\ \sum_{\alpha,\beta,\gamma,\delta}
  s_\alpha(x)\ s_\beta(x)\
  s_{\{1^2\}\otp\gamma}(x)\ s_{\{2\}\otp\delta}(x)\cr
&\qquad\qquad\qquad\qquad\cdot
s_{\{1^2\}\otp\alpha}(y)\ s_{\{2\}\otp\beta}(y)\ s_\gamma(y)\ s_\delta(y),
\label{Eq-M21xy} \\
M_{1^3}(x,y)
&=M_{1^3}\,(x)M_{1^3}(y)\ \sum_{\sigma,\tau} 
  s_{\sigma}(x)\ s_{\{1^2\}\otp\tau}(x)\cdot 
  s_{\{1^2\}\otp\sigma}(y)\ s_\tau(y),
\label{Eq-M111xy} \\
M_{1^4}(x,y)
&=M_{1^4}(x)\,M_{1^4}(y)
  \sum_{\rho,\sigma,\tau} 
  s_{\rho}(x)\ s_{\{1^2\}\otp\sigma}(x)\ s_{\{1^3\}\otp\tau}(x)\nn
&\qquad\qquad\qquad\qquad\cdot
  s_{\{1^3\}\otp\rho}(y)\ s_{\{1^2\}\otp\sigma}(y)\ s_\tau(y).
\label{Eq-M1111xy}
\end{align}
\myeenv

\subsection{Series quotients of Schur function products}

The previous results allow us to prove the following:
\mybenv{Proposition}\label{Prop-skewMpi}
For any partitions $\pi$, $\mu$ and $\nu$
\begin{align}
(\{\mu\}\,\{\nu\})/M_\pi
&\ =\,  
\sum_{\sigma(\xi,\eta,k)}\ 
  \{\mu/\big(M_\pi\!\!
\prod_{\xi,\eta<\pi}\ 
  \prod_{k=1}^{C^\pi_{\xi\eta}}
  \xi\otp\sigma(\xi,\eta,k)\big)\}
\ \
  \{\nu/\big(M_\pi\!\!
\prod_{\xi,\eta<\pi}\ 
  \prod_{k=1}^{C^\pi_{\xi\eta}}
  \eta\otp\sigma(\xi,\eta,k)\big)\}\,,
\label{Eq-skewMpi}
\end{align}
where for each $\xi$, $\eta$ and $k$ the summation is
carried out over all partitions $\sigma(\xi,\eta,k)$.
\myeenv

This proposition will be a corollary to our Main Theorem, so we postpone
the proof. The reader who feels uneasy with our Hopf algebraic proof might
like to compare the proof of proposition~\ref{Prop-skewMpi} by
combinatorial means in appendix \ref{sec:combProofs} and is invited to
compare with the further development using Hopf algebras.  It should be
noted in particular that the Lemmas~\ref{Lem-opskew} and \ref{Lem-opZskew}
becoming implicit in the structural definitions of the Hopf algebraic
machinery rather than requiring separate proofs.

\mybenv{Example}
Proposition~\ref{Prop-skewMpi} encompasses by way of example the following results:
\begin{align}
 (\{\mu\}\,\{\nu\})/M
&=\{\mu/M\}\,\{\nu/M\},
\label{Eq-skewM} \\
 (\{\mu\}\,\{\nu\})/M_2
&=\sum_{\sigma} 
 \{\mu/(M_2\,\sigma)\}\, 
 \{\nu/(M_2\,\sigma)\},
\label{Eq-skewM2} \\
 (\{\mu\}\,\{\nu\})/M_{1^2}
&=\sum_{\sigma} 
 \{\mu/(M_{1^2}\,\sigma)\}\ 
 \{\nu/(M_{1^2}\,\sigma)\},
\label{Eq-skewM11} \\
 (\{\mu\}\,\{\nu\})/M_3
&=\sum_{\sigma,\tau} 
 \{\mu/(M_3\,\sigma\,(2\otp\tau))\}\, 
 \{\nu/(M_3\,(2\otp\sigma)\,\tau)\},
\label{Eq-skewM3} \\
 (\{\mu\}\,\{\nu\})/M_{21}
&=\sum_{\alpha,\beta,\gamma,\delta} 
 \{\mu/(M_{21}\,\alpha\,\beta\,(1^2\otp\gamma)\,(2\otp\delta))\}
\nn
&\hskip3cm
 \{\nu/(M_{21}\,(1^2\otp\alpha)\,(2\otp\beta)\,\gamma\,\delta)\},
\label{Eq-skewM21} \\
 (\{\mu\}\,\{\nu\})/M_{1^3}
&=\sum_{\sigma,\tau} 
 \{\mu/(M_{1^3}\,\sigma\,(1^2\otp\tau))\}\ 
 \{\nu/(M_{1^3}\,(1^2\otp\sigma)\,\tau)\},
\label{Eq-skewM111} \\
 (\{\mu\}\,\{\nu\})/M_{1^4}
&=\sum_{\rho,\sigma,\tau} 
 \{\mu/(M_{1^4}\,\rho\,(1^2\otp\sigma)\,(1^3\otp\tau))\}\nn
&\hskip3cm
 \{\nu/(M_{1^4}\,(1^3\otp\rho)\,(1^2\otp\sigma)\,\tau)\}.
\label{Eq-skewM1111} 
\end{align}
\myeenv

\subsection{$\pi$-Newell-Littlewood product theorem}

Having identified formal characters of representations of subgroups
$\grpH_\pi(n)$ of $\grpGL(n)$ in (\ref{Eq-brinvHpi}), we are in a  position
to combinatorially decompose their tensor products. This may be done by
writing such products as products of Schur functions using 
(\ref{Eq-brinvHpi}), evaluating the Schur function products by means of
(\ref{Eq-outerLR}) and then restricting the corresponding characters of
$\grpGL(n)$ to its subgroup $\grpH_\pi(n)$ by means of (\ref{Eq-brHpi}).
Our previous Propositions allow us to simplify the results and obtain the
general formula given in \medskip

\mybenv{Proposition}
\label{Prop-pHpi}
{Let $\(\mu\)$ and $\(\nu\)$ be formal characters of $\grpH_\pi$. 
Then 
\begin{align}
\(\mu\)\,\(\nu\)
&=
\sum_{\sigma(\xi,\eta,k)}
  \(\{\mu/\prod_{\xi,\eta<\pi} 
  \prod_{k=1}^{C^\pi_{\xi\eta}} \{\xi\}\otp\sigma(\xi,\eta,k)\}\,
  \{\nu/\prod_{\xi,\eta<\pi} 
  \prod_{k=1}^{C^\pi_{\xi\eta}} \{\eta\}\otp\sigma(\xi,\eta,k)\}\),
\label{Eq-pHpi}
\end{align}
where for each $\xi$, $\eta$ and $k$ the summation is carried out 
over all partitions $\sigma(\xi,\eta,k)$.}
\myeenv

The combinatorial proof is given in appendix \ref{sec:combProofs}. The
theorem will be an easy consequence of our Main Theorem in section
\ref{sec:Hopf}.

\mybenv{Example}
Once again Proposition~\ref{Prop-pHpi} may be illustrated by means of 
examples:
\begin{align}
\hskip-0.5cm
\grpH_1(n)=\grpGL(n\!\!-\!\!1)&:&&&  \{\mu\} \ \{\nu\}
&=\{(\mu)\,(\nu)\},
\label{Eq-pH1} \\
\hskip-0.5cm
\grpH_2(n)=\grpO(n)&:&&&  [\mu]\ [\nu]
&=\sum_{\sigma} 
 [(\mu/\sigma)(\nu/\sigma)],
\label{Eq-pH2} \\
\hskip-0.5cm
\grpH_{1^2}(n)=\grpSp(n)&:&&&  \la\mu\ra\ \la\nu\ra
&=\sum_{\sigma,\tau} 
 \la(\mu/\sigma)(\nu/\tau)\ra,
\label{Eq-pH11} \\
\hskip-0.5cm
\grpH_{3}(n)&:&&&  \(\mu\)\ \(\nu\)
&=\sum_{\sigma,\tau} 
 \(\,(\mu/\sigma\,(2\otimes\tau))\, 
    (\nu/(2\otimes\sigma)\,\tau)\,\),
\label{Eq-pH3} \\
\hskip-0.5cm
\grpH_{21}(n)&:&&&  \(\mu\)\ \(\nu\)
&=\sum_{\alpha,\beta,\gamma,\delta} 
 \(\, (\mu/\alpha\,\beta\,(1^2\otimes\gamma)\,(2\otimes\delta))\,
\cr
\hskip-0.5cm
&&&&&\quad\quad\cdot
      (\nu/(1^2\otimes\alpha)\,(2\otimes\beta)\,\gamma\,\delta)\, \),
\label{Eq-pH21} \\
\hskip-0.5cm
\grpH_{1^3}(n)&:&&&  \(\mu\)\ \(\nu\)
&=\sum_{\sigma,\tau} 
 \(\, (\mu/\sigma\,(1^2\otimes\tau))\,
      (\nu/(1^2\otimes\sigma)\,\tau)\, \),
\label{Eq-pH111} \\
\hskip-0.5cm
\grpH_{1^4}(n)&:&&&  \(\mu\)\ \(\nu\)
&=\sum_{\rho,\sigma,\tau} 
 \(\, (\mu/\rho\,(1^2\otimes\sigma)\,(1^3\otimes\tau))\,
\cr
\hskip-0.5cm
&&&&&\quad\quad\cdot
      (\nu/(1^3\otimes\rho)\,(1^2\otimes\sigma)\,\tau) \,\).
\label{Eq-pH1111} 
\end{align}
\myeenv

\section{\label{sec:Hpi}Nature of the non-classical groups $\grpH_\pi$}
\setcounter{equation}{0}

Before developing the technicalities of our machinery in section
\ref{sec:Hopf}, we try in the present section to give some hints as to what kinds
of groups $\grpH_\pi$ are to be expected. The subgroups $\grpH_\pi(n)$ of $\grpGL(n)$ leaving invariant a fixed
tensor of Young symmetry type $\pi$ are not necessarily reductive, let
alone semi simple Lie groups. The characters $\(\lambda\)_{\pi}$ 
which we study correspond to representations which may be reducible, 
but not necessarily fully reducible, for $|\pi|>2$. The full resolution
of these issues is beyond the scope of the present introductory study. 

In dealing with concrete examples we
have to pass from formal characters to actual characters which implies that
there are syzygies, and the representations are not in general free modules.
Thus for each
$\pi$ we expect a set of `standard' characters $\(\lambda\)_{\pi}$,
together with so-called `modification rules'  for non-standard ones
\cite{king:1971a,SCHUR}. As mentioned already, a systematic treatment  is beyond the scope of
the paper, but some hints can be inferred from working through the examples. At the present level of discussion
modification rules must be established on a case-by-case basis, and may even depend for each
$\pi$ on different canonical forms of the invariant tensor of symmetry type
$\pi$. For present purposes we simply regard the  $\(\lambda\)_{\pi}$ as a
list of formal characters associated with the  group $\grpH_\pi$. Here and in
the  following subsections, we take up the  case $\pi = \{ 1^{3} \}$, and
discuss some details of branching, product  and modification rules for the
formal characters $\(\lambda \)_{1^{3}}$.  We drop the index $\pi=\{1^3\}$
from now on for brevity and notational  clarity. In appendix \ref{Tables}
below further branching and product formulae are given for other cases
$\{3\}$, $\{ 21\}$ but without any analysis of  modification rules.

\subsection{$\grpH_{1^3}(n)$ in dimension $n=3$}

\subsubsection{$\grpSL(3)\equiv\grpH_{1^3}(3)$}

Consider the case $\pi = \{1^{3}\}$ corresponding to the
existence of an invariant totally antisymmetric third rank tensor
$\eta_{ijk}$ satisfying the conditions:
\begin{align}
\eta_{ijk}
&=\eta_{jki}=\eta_{kij}=-\eta_{ikj}=-\eta_{jik}=-\eta_{kji}
\label{Eq-eta}
\end{align}
The requirement that this tensor be invariant under the action of all 
elements $A\in H_{1^3}(n)$ gives rise to the constraints
\begin{align}
A&:\ \eta_{ijk} \rightarrow A_i^p\,A_j^q\,A_k^r\ \eta_{pqr}\ =\ \eta_{ijk}
\label{Eq-etaA}
\end{align}
with $i,j,k,p,q,r\in\{1,2,\ldots,n\}$. Here and in what follows 
Einstein's convention is followed whereby repeated indices, such as 
$p$, $q$ and $r$, are summed over their full range of values, 
in this case $1,2,\ldots,n$.

In fact for given $n$ there may be more than one canonical form of the
invariant tensor $\eta_{ijk}$. For $n=3$ the canonical form is  necessarily
defined by $\eta_{ijk}=a\epsilon_{ijk}$ with $\epsilon_{ijk}$ the usual
third rank totally antisymmetric tensor  in a three-dimensional space such
that $\epsilon_{123}=1$. As can be seen from the constraint conditions
(\ref{Eq-etaA}) we can take $a=1$  without loss of generality and set
$\eta_{ijk}=\epsilon_{ijk}$. Using this in (\ref{Eq-etaA}) with $n=3$ just
gives $\eta_{123}=\det A=1$ and up to isomorphism we can immediately make
the identification $\grpH_{1^3}(3)=\grpSL(3)$. In this case, therefore, the
relevant subgroup of $\grpGL(3)$ is the semisimple  Lie group $\grpSL(3)$.
All  its representations are fully reducible and we can identify the 
characters $\(\lambda\)$ with Schur functions $s_\lambda(x_1,x_2,x_3)$ 
with constraint $x_1x_2x_3=1$ for partitions $\lambda$ of length less than
$3$. For other cases certain modification rules are required, to  which we
shall return later. It is immediate that a similar analysis can be carried
out for any $\grpSL(n)=\grpH_{1^n}(n)$, which gives a unified character
theory for all $\grpSL(n)$ groups.

\subsubsection{$\grpH_{1^3}(3)$: Modification rules and 
$\grpGL(3)\supset H_{1^3}(3)=\grpSL(3)$ branching rules}

This special case should be trivial because it corresponds  to the well
known restriction from $\grpGL(3)$ to $\grpSL(3)$. However, even here
although the branchings are indeed trivial, the modification rules are 
somewhat  complicated in our formalism.

The branching rule from characters $\{\lambda\}$ of $\grpGL(3)$
to formal characters $\(\mu\)$ of $\grpH_{1^3}(3)=\grpSL(3)$ takes 
the form:
\begin{align}
\grpGL(3)\supset \grpH_{1^3}(3):
&&& \{\lambda\}\rightarrow\(\lambda/M_{1^3}\)
\end{align}
We use the double parentheses $\(\lambda\)_\pi$ for the formal  characters
of $\grpH_{\pi}$. If the context is clear we drop the index, which may be
replaced by the dimension of the representation. Confusion between integer
dimension and partitions cannot  occur. The particular series employed here
reads
\begin{align}
M_{1^3}
&=
\{1^3\}\otp M =  \{0\} + \{1^3\} + \{2^3\} + \{3^3\} + \{4^3\} + \cdots , 
\end{align}
where it has only been necessary to retain terms of length
no greater than $3$. This yields the following formal characters, 
where the subscripts give the dimension of the corresponding
representations, or more precisely the value of the characters
and formal characters at the identity.
\begin{align}
\begin{array}{|l|l|}
\hline
\{\lambda\}_{\textrm{dim}}&\(\lambda/M_{1^3}\)_{\textrm{dim}}\\
\hline
\{0\}_{1}&\(0\)_{1}\\
\{1\}_{3}&\(1\)_{3}\\
\{11\}_{3}&\(11\)_{3}\\
\{111\}_{1}&\(111\)_{0}+\(0\)_{1}\\
\{1111\}_{0}&\(1111\)_{-3}+\(1\)_{3}\\
\{2\}_{6}&\(2\)_{6}\\
\{21\}_{8}&\(21\)_{8}\\
\{211\}_{3}&\(211\)_{0}+\(1\)_{3}\\
\{2111\}_{0}&\(2111\)_{-9}+\(2\)_{6}+\(11\)_{3}\\
\{22\}_{6}&\(22\)_{6}\\
\{221\}_{3}&\(221\)_{0}+\(11\)_{3}\\
\{2211\}_{0}&\(2211\)_{-8}+\(21\)_{8}+\(111\)_{0}\\
\{222\}_{1}&\(222\)_{0}+\(111\)_{0}+\(0\)_{1}\\
\{2221\}_{0}&\(2221\)_{0}+\(211\)_{0}+\(1111\)_{-3}+\(1\)_{3}\\
\{2222\}_{0}&\(2222\)_{3}+\(2111\)_{-9}+\(2\)_{6}\\
\{3\}_{10}&\(3\)_{10}\\
\{31\}_{15}&\(31\)_{15}\\
\{311\}_{6}&\(311\)_{0}+\(2\)_{6}\\
\{3111\}_{0}&\(3111\)_{-18}+\(3\)_{10}+\(21\)_{8}\\
\{32\}_{15}&\(32\)_{15}\\
\hline
\end{array}
\end{align}
For $\ell(\mu)\leq 2$, as expected for the branching from $\grpGL(3)$ to
$\grpH_{1^3}(3)=\grpSL(3)$, we have  
$\{\lambda_1,\lambda_2\}\rightarrow\(\lambda_1,\lambda_2\)$. On the other
hand for $\ell(\mu)\geq3$ we require modification rules to interpret the
formal characters of $\grpH_{1^3}(n)=\grpSL(3)$  in terms of irreducible
characters. From the above branchings these must include the following:
\begin{align}
\(111\)_{0}&=0 \nn
\(211\)_{0}&=0 \nn
\(221\)_{0}&=0 \nn
\(222\)_{0}&=0 \nn
\(311\)_{0}&=0 \nn
\(1111\)_{-3} &=-\(1\)_{3} \nn
\(2111\)_{-9}&=-\(2\)_{6}-\(11\)_{3} \nn
\(2211\)_{-8}&=-\(21\)_{8} \nn
\(2221\)_{0} &= 0 \nn
\(2222\)_{3}  &=\(11\)_{3} \nn
\(3111\)_{-18}&=-\(3\)_{10}-\(21\)_{8} \nn
\end{align}
More generally, for any $\(\lambda_1,\lambda_2,\lambda_3,\lambda_4\)$ 
with $\lambda_1\geq\lambda_2\geq\lambda_3\geq\lambda_4\geq0$ 
the following constitute a complete set of modification rules:
\begin{align}
\(\lambda_1,\lambda_2,\lambda_3,0\)&=0 
     \ \ \hbox{if}\ \lambda_3\geq1 \nn
\(\lambda_1,\lambda_1,1,1\)&=-\(\lambda_1,\lambda_1-1\)
     \ \ \hbox{if}\ \lambda_1=\lambda_2\geq1 \nn
\(\lambda_1,\lambda_2,1,1\)&=-\(\lambda_1,\lambda_2-1\)
                              -\(\lambda_1-1,\lambda_2\) 
     \ \ \hbox{if}\ \lambda_1>\lambda_2\geq1 \nn
\(\lambda_1,\lambda_2,2,1\)&=0
     \ \ \hbox{if}\ \lambda_1\geq\lambda_2\geq2 \nn
\(\lambda_1,\lambda_2,2,2\)&=\(\lambda_1-1,\lambda_2-1\)
     \ \ \hbox{if}\ \lambda_1\geq\lambda_2\geq2 \nn
\(\lambda_1,\lambda_2,\lambda_3,\lambda_4\)&=0
     \ \ \hbox{if}\ \lambda_3\geq3 
\end{align}

\subsubsection{Product of characters}

As a check we can recover the known $\grpSL(3)$ products through the use 
of
\begin{align}
\(\mu\) \, \(\nu\) &= \sum_{\alpha,\beta}
     \(\,(\mu/\alpha\cdot\{(1^2)\}\otp\beta)\cdot 
         (\nu/\{(1^2)\}\otp\alpha\}\cdot\beta)\,\).
\end{align}
For example,
\begin{align}
\(22\)_6\, \(21\)_8&= 
  \( (22)\cdot(21) \)+ \( (22/1)\cdot(21/11) \) 
  +  \( (22/11)\cdot(21/1) \) \nn
&+  \( (22/(1\cdot11))\cdot(21/(11\cdot1))\) 
 +  \( (22/22)\cdot(21/2) \)\nn
&=\ \(43\)+\(421\)+\(331\)+\(322\)+\(3211\)+\(2221\)\nn
& + 2\(31\)+2\(22\)+3\(211\)+\(1111\)+2\(1\)\nn
&= \(43\)_{24}+\(31\)_{15}+\(22\)_{6}+\(1\)_{3}
\end{align}
where the modification rules have been used in the last step. The result
agrees with what we obtain by going up (trivially) from
$\grpH_{1^3}(3)=\grpSL(3)$ to $\grpGL(3)$, carrying out the product  in
$\grpGL(3)$ and branching to $\grpSL(3)$ by throwing away columns of length
greater than $3$.
\begin{align}
 \(22\)_6\, \(21\)_8&= \{22\}\, \{21\} = \{43\}+\{421\}+\{331\}+\{322\}\nn
   &= \{43\}+\{31\}+\{22\}+\{1\} \nn
   &= \(43\)_{24}+\(31\)_{15}+\(22\)_{6}+\(1\)_{3}.
\end{align}
 
\subsection{$\grpH_{1^3}(4):$ branching rules, modification
rules, and products of characters.}

\subsubsection{Matrix realization}

Turning to the case $n=4$ the canonical form of $\eta$ is such that
with a suitable scaling 
\begin{align}
\eta_{ijk}
&=
\begin{cases} 
\epsilon_{abc} 
& \text{for~} (i,j,k)=(a,b,c) \text{~with~} a,b,c\in\{1,2,3\}; \cr
0 
& \text{otherwise.} \cr 
\end{cases}
 \label{Eq-eta4}
\end{align}
Thus the constraints (\ref{Eq-etaA}) reduce to
\begin{align}
A_i^a\,A_j^b\,A_k^c\,\epsilon_{abc}&=\eta_{ijk},
\label{Eq-Aabc}
\end{align}
so that
\begin{align}
 &A_4^a\,A_4^b\,A_4^c\,\epsilon_{abc}=\eta_{444}=0,
\label{Eq-A444} \\
 &A_p^a\,A_4^b\,A_4^c\,\epsilon_{abc}=\eta_{p44}=0,
\label{Eq-Ap44} \\
 &A_p^a\,A_q^b\,A_4^c\,\epsilon_{abc}=\eta_{pq4}=0,
\label{Eq-Apq4} \\
 &A_p^a\,A_q^b\,A_r^c\,\epsilon_{abc}=\eta_{pqr}=\epsilon_{pqr},
\label{Eq-Apqr} 
\end{align}
with $p,q,r\in\{1,2,3\}$. The first two constraints 
(\ref{Eq-A444}) and (\ref{Eq-Ap44}) are satisfied
automatically because of the antisymmetry of $\epsilon_{abc}$.
The fourth constraint (\ref{Eq-Apqr})
gives $\det B=1$ where $B$ is the $3\times3$
submatrix of $A$ such that $B_a^b=A_a^b$ for all $a,b\in\{1,2,3\}$.
With this notation (\ref{Eq-Apqr}) becomes
\begin{align}
B_p^a\,B_q^b\,B_r^c\,\epsilon_{abc}=\epsilon_{pqr},
\label{Eq-Bpqr}
\end{align}
so that 
\begin{align}
{B^{-1}}_s^r \epsilon_{pqr} = {B_p}^a \, {B_q}^b \, {B^{-1}}_s^r {B_r}^c \,\epsilon_{abc}
    = {B_p}^a \, {B_q}^b \, \epsilon_{abs}.
\label{Eq-Binv}
\end{align}
Using this in (\ref{Eq-Apq4}) then gives
\begin{align}
A_4^c\,{B^{-1}}_c^r\epsilon_{pqr}=0.
\label{Eq-ABinv}
\end{align}
Since this is true for all $p,q\in\{1,2,3\}$ it follows that
\begin{align}
A_4^c\,{B^{-1}}_c^r=0
\label{Eq-A4Br}
\end{align}
for all $r\in\{1,2,3\}$. Hence
\begin{align}
 A_4^c\,{B^{-1}}_c^r\,B_r^d=A_4^d=0
\label{Eq-ABBA}
\end{align}
for all $d\in\{1,2,3\}$.
Thus $A$ necessarily takes the form
\begin{align}
A
&=
\begin{pmatrix}B&D\cr0&C\end{pmatrix}
\quad\hbox{with}\quad \det B=1,\ \det C\neq0.
\label{Eq-A}
\end{align}
where $D$ is an arbitrary $3\times1$ matrix, $C$ is a non-zero  $1\times1$
matrix and $0$ signifies a $1\times3$ zero matrix in accordance with
(\ref{Eq-ABBA}). It follows that $\grpH_{1^3}(4)$ is subgroup of
$\grpGL(4)$ consisting of non-singular $4\times4$ matrices of the form
(\ref{Eq-A}). It should be noted that this is an example of an affine
group.  It is neither semisimple nor reductive, as can be seen  from the
block triangular form of the defining $4$-dimensional reducible, but
indecomposable, representation (\ref{Eq-A}). It contains the reductive
group $\grpSL(3)\times \grpGL(1)$ as  a proper subgroup. In fact, we have
shown that the formal character theory  can be extended to non-semisimple
groups in a straight forward  manner. The presented example
$\grpH_{1^3}(4)$ contains after  the further branching
$\grpSL(3)\downarrow\grpSO(3)$ the group of  motions of a rigid body and
has as such potential applications in robotics etc.

\subsubsection{Branching rules}

First note that the inequivalent finite-dimensional rational  irreducible
representations of $\grpGL(4)$ have characters $\varepsilon^r
\{\lambda\}=\varepsilon^r s_\lambda(x)$  with $r$ an integer, and
$\lambda=(\lambda_1,\lambda_2,\lambda_3,\lambda_4)$ a partition of length
$\ell(\lambda)\leq 4$. The parameters $x=(x_1,x_2,x_3,x_4)$ are the
eigenvalues of  the group elements $A\in GL(4)$, and $\varepsilon$ is the
character of the representation $A\mapsto\det(A)$, so that
$\varepsilon=s_{1^4}(x)=\{1^4\}=x_1x_2x_3x_4$. 

The branching rule from characters $\{\lambda\}$ of $\grpGL(4)$ to formal
characters $\(\mu\)$ of $\grpH_{1^3}(4)$ takes the form:
\begin{align}
\grpGL(4)\supset \grpH_{1^3}(4)&:&&  
\{\lambda\}\rightarrow\(\lambda/M_{1^3}\)
\end{align} 
with
\begin{align}
M_{1^3} &= \{ 1^3 \}\otp M =  {\{} 0\} + {\{} 1^3\} + {\{}2^3\}
    + \{3^3\} + \{4^3\} + \cdots , 
\end{align}
where it has only been necessary to retain terms of length no greater than
$4$. This yields the following data, where the subscripts give the
dimension of the corresponding representations, or more precisely the value
of the characters and formal characters at the identity.
\begin{align}
\begin{array}{|l|l|}
\hline
\{\lambda\}_{\textrm{dim}}&\(\lambda/M_{1^3}\)_{\textrm{dim}}\\
\hline
\{0\}_{1}&\(0\)_{1}\\
\{1\}_{4}&\(1\)_{4}\\
\{11\}_{6}&\(11\)_{6}\\
\{111\}_{4}&\(111\)_{3}+\(0\)_{1}\\
\{1111\}_{1}&\(1111\)_{-3}+\(1\)_{4}\\
\{2\}_{10}&\(2\)_{10}\\
\{21\}_{20}&\(21\)_{20}\\
\{211\}_{15}&\(211\)_{11}+\(1\)_{4}\\
\{2111\}_{4}&\(2111\)_{-12}+\(2\)_{10}+\(11\)_{6}\\
\{22\}_{20}&\(22\)_{20}\\
\{221\}_{20}&\(221\)_{14}+\(11\)_{6}\\
\{2211\}_{6}&\(2211\)_{-17}+\(21\)_{20}+\(111\)_{3}\\
\{222\}_{10}&\(222\)_{6}+\(111\)_{3}+\(0\)_{1}\\
\{2221\}_{4}&\(2221\)_{-8}+\(211\)_{11}+\(1111\)_{-3}+\(1\)_{4}\\
\{2222\}_{1}&\(2222\)_{3}+\(2111\)_{-12}+\(2\)_{10}\\
\{3\}_{20}&\(3\)_{20}\\
\{31\}_{45}&\(31\)_{45}\\
\{311\}_{36}&\(311\)_{26}+\(2\)_{10}\\
\{3111\}_{10}&\(3111\)_{-30}+\(3\)_{20}+\(21\)_{20}\\
\{32\}_{60}&\(32\)_{60}\\
\hline
\end{array}
\end{align}

\subsubsection{Aspects of $\grpH_{1^3}(4)$ modification rules}

To interpret the formal characters of $\grpH_{1^3}(4)$ it is 
necessary to apply modification rules. These are required for
all $\(\mu\)$ with $\mu$ of length $\ell(\mu)=4$. In 
these branchings, these formal characters can only arise
in cases for which $\lambda$ also has length $4$. However,
as pointed out above $\{1^4\}=\varepsilon$ is the 
character of the $1$-dimensional determinant representation
of $\grpGL(4)$. It follows that
\begin{align}
\{\lambda_1,\lambda_2,\lambda_3,\lambda_4\}=\varepsilon^{\lambda_4}\
\{\lambda_1-\lambda_4,\lambda_2-\lambda_4,\lambda_3-\lambda_4,0\}.
\end{align}
Applying this to the tabulated identities gives:
\begin{align}
\(1111\)_{-3} &=\{1^4\}_{1}-\(1\)_{4} \nn
              &=\varepsilon\(0\)_1-\(1\)_{4} \nn
\(2111\)_{-12}&=\{2111\}_{4}-\(2\)_{10}-\(11\)_{6} \nn
              &=\varepsilon\(1\)_{4}-\(2\)_{10}-\(11\)_{6} \nn
\(2211\)_{-17}&=\{2211\}_{6}-\(21\)_{20}-\(111\)_{3} \nn
              &=\varepsilon\(11\)_{6}-\(21\)_{20}-\(111\)_{3} \nn
\(2221\)_{-8} &=\{2221\}_{4}-\(211\)_{11}-\(1111\)_{-3}-\(1\)_{4} \nn
              &=\varepsilon\{111\}_{4}-\(211\)_{11}
                   -\varepsilon\(0\)_1+\(1\)_{4}-\(1\)_{4} \nn
              &=\varepsilon( \(111\)_{3}+\(0\)_{1} )-\(211\)_{11}
                   -\varepsilon\(0\)_1 \nn
              &=\varepsilon\(111\)_{3}-\(211\)_{11} \nn
\(2222\)_{3}  &=\{2222\}_1-((2111\)_{-12}-\(2\)_{10} \nn
              &=\varepsilon^2\{0\}_1-\varepsilon\(1\)_{4}+\(2\)_{10}
                    +\(11\)_{6}-\(2\)_{10} \nn
              &=\varepsilon^2\(0\)_1-\varepsilon\(1\)_{4}+\(11\)_{6} \nn
\(3111\)_{-30}&=\{3111\}_{10}-\(3\)_{20}-\(21\)_{20} \nn
              &=\varepsilon\(2\)_{10}-\(3\)_{20}-\(21\)_{20}
\end{align}
This gives a collection of modification rules to be applied in the
case $\(\mu\)$ with $\ell(\mu)=4$. Of course, one would like to have a complete set of modification
rules, including those appropriate to $\(\mu\)$ with $\ell(\mu)>4$.

\subsubsection{Product of characters}

We can exploit this to analyse products of characters. For example,
\begin{align}
&\(2\)_{10}\, \(111\)_{3} 
  = \(311\)_{26}+\(2111\)_{-12}+\(2\)_{10}+\(11\)_{6}\nn
&\ = \(311\)_{26}+(\varepsilon\(1\)_{4}-\(2\)_{10}-\(11\)_{6})
         +\(2\)_{10}+\(11\)_{6}\nn
&\ = \(311\)_{26}+\varepsilon\(1\)_{4}
\end{align}
Further examples of products of $\grpH_{1^3}(4)$ characters are given in appendix
\ref{subsec:grpH_1^3}.

Clearly by proceeding in this way we can build up branching rules, modification rules and product rules for characters for each 
$\grpH_{\pi}$ on a case-by-case basis (including for example 
obtaining each $\grpSL(n)$ as  a subgroup $\grpH_{1^n}(n)$ of $\grpGL(n)$ in a unified
framework). However, we now close this heuristic section
and pause for  an introduction to Hopf algebra methods, so as  to
start to develop these results into a general theory.

\section{\label{sec:Hopf}Hopf algebraic analysis}
\setcounter{equation}{0}

In section \ref{sec:SSeries} the series $M_{\pi} = \{\pi\} \otp M$ and 
associated formal characters for non-classical subgroups were derived from 
straightforward generalisations of the known combinatorial route to the 
$\grpGL(n)$ subgroups $\grpGL(n-1)$, $\grpO(n)$ and $\grpSp(n)$. The aim 
of the present section is to embed the combinatorial proofs, given above 
and in the appendix, into the discussion of branching and product rules,
in the context of the Hopf algebraic structure of, and the associated
cohomological framework for, symmetric functions as developed in
\cite{fauser:jarvis:2003a}. The Hopf algebraic aspects are well-known in
the combinatorial literature \cite{geissinger:1977a,zelevinsky:1981a,%
zelevinsky:1981b,thibon:1991a,thibon:1991b} 

In \cite{fauser:jarvis:2003a} these branching rules were studied from the
point of view of cliffordization (multiplicative deformations) of the
algebraic structure, and initial steps were taken towards cohomological
classifications. Here and in section \ref{subsec:HopfSeries} below  we give
in outline the basic setting needed to clarify the relationship between
series-induced branching rules, and associative deformations  (2-cocycles).
A detailed development
of the Hopf algebra cohomology involved is given in
\cite{brouder:fauser:frabetti:oeckl:2002a}.  This
algebraic structure is used in a similar manner in \cite{fauser:jarvis:2003a}, and will be
utilized in \cite{fauser:jarvis:2005a}. In section
\ref{subsec:HopfNewBranchings}, standard rules for
plethysm distributivity are introduced, transcribed into the language of products  and
coproducts of symmetric functions. Finally section \ref{subsec:Kernels}
leads into the presentation of a very general class of 2-cochains, of which
the present $\pi$-branching rules provide examples which are also 
2-cocycles. The abstract development is complemented in the appendix 
\ref{sec:GraphCalc} by a brief introduction to a systematic diagrammatical
approach using tangles.

The basic structure of interest is the outer Hopf  algebra of the  ring of
symmetric functions $\Lambda$. Given the canonical Schur scalar product
$\langle \cdot \mid \cdot \rangle : \Lambda \otimes \Lambda  \rightarrow {\openZ}$ which
makes the Schur functions orthonormal,
\begin{align}
    \langle s_{\lambda}\mid s_{\mu}\rangle =&\, \delta_{\lambda \mu},
\end{align}
we define the outer coproduct of any symmetric function $F$,
$\Delta(F)$, by duality, that is
\begin{align}
    \langle\Delta(F)\mid G \otimes H\rangle := &\, \langle F\mid G \cdot H \rangle
\end{align}
for any symmetric functions $G$ and $H$, where $\cdot$ is outer 
multiplication. In the basis of Schur functions this yields using 
(\ref{Eq-coLR})
\begin{align}
    \Delta(s_{\pi}) = &\,  \sum_{\xi, \eta} C^\pi_{\xi \eta} 
    s_{\xi}\otimes s_{\eta}.
\end{align}
There is a similar inner coproduct which dualizes inner multiplication 
of symmetric functions,
\begin{align}
    \langle\delta(F)\mid G \otimes H\rangle := &\, \langle F\mid G \ip H \rangle
\end{align}
for which the structure constants in the Schur function basis are 
therefore the structure constants of inner multiplication. In both 
cases we can adapt the Sweedler convention of affixing bracketed 
subscripts to denote the list of parts of the coproducts. To distinguish
different coproducts we employ the Brouder-Schmitt convention
\cite{brouder:schmitt:2002a} using (1), (2) for the outer coproduct
$\Delta$ and $[1]$, $[2]$ for the inner coproduct $\delta$:
\begin{align}
\Delta(F) &= \sum_{(F)} F_{(1)}\otimes F_{(2)} = F_{(1)}\otimes F_{(2)} \nn
\delta(F) &= \sum_{[F]} F_{[1]}\otimes F_{[2]} = F_{[1]}\otimes F_{[2]}. 
\end{align}
For later use we define \textit{proper cuts} of a coproduct to be the sum 
of those tensors which do not contain the unit. The proper cut part is 
denoted by primes,
\begin{align}\label{PK}
\Delta^\prime(F)&=F^\prime_{(1)}\otimes F^\prime_{(2)}
\,=\, \Delta(F) -F\otimes 1 - 1\otimes F.
\end{align}
The remaining ingredients of the outer Hopf algebra are the antipode
antihomomorphism map 
\begin{align}
{\sf S}(\{\mu \}) &= (-1)^{|\mu|}\{\mu^\prime \},
\label{Eq-antipode}
\end{align}
where $\mu^\prime$ is the partition conjugate to $\mu$, and the counit
character is given by
\begin{align} 
\epsilon(\{\mu\}) &= \delta_{\mu 0}.
\label{Eq-counit}
\end{align}
The inner product and inner coproduct do not form a Hopf algebra, but only 
inner convolution algebras. The reader interested in the details may
consult \cite{fauser:jarvis:2003a}.

\subsection{Series-induced branching operators, cliffordization and cohomology}
\label{subsec:HopfSeries}

Branchings have a very neat description in terms of Schur function series.
One can use Hopf algebra methods to encode this. A linear form $\phi :
\Lambda\rightarrow \openZ$ acting on the ring of symmetric functions 
$\Lambda$ with values in $\openZ$ is called a 1-cochain. Using such 
1-cochains $\phi$, the branching operator obtained by skewing with a 
series $/\Phi$ reads in Hopf algebraic terms
\begin{align}\label{brOp}
s_{\lambda/\Phi} &= \{\lambda\}/\Phi = (\phi\otimes \Id)\Delta(s_\lambda)
\end{align}
where for given $\Phi$ the 1-cochain $\phi$ is defined by 
\begin{align}\label{phi}
\phi(s_\mu)=\langle \Phi\,|\,s_\mu\rangle
\end{align} 
The operator $/\Phi$ is called branching operator, recalling the group
branchings of (\ref{Eq-brGLn-1},\ref{Eq-brOn},\ref{Eq-brSpn}). We adopt
the convention that 1-cochains are denoted by lower case letters and the
corresponding Schur function series by upper case letters. Given a
$\grpGL(n)$ character a branching process using an  appropriate branching
operator yields Schur functions describing the formal characters of the
appropriate $\grpGL(n)$ subgroup. Under the convolution product one finds
an inverse series $\Phi^{-1}$  which allows one to undo the branching. The
branching operator $/\Phi$ and its convolutive inverse $/\Phi^{-1}$ can be
used to define a new product
\begin{align}\label{prodPhi}
\(A\) \cdot_\phi \(B\) 
&=  \(\,(A/\Phi^{-1}\, \cdot\, B/\Phi^{-1})/\Phi\).
\end{align}
It is remarkable, that one can derive a 2-cocycle $(\partial\phi)$ from 
the linear form $\phi$, which allows one to reinterpret this process as a
cliffordization or twist. Using Sweedler indices  (see above)  we can
rewrite the deformed product as
\begin{align}
A \cdot_\phi B 
&=  (\partial\phi)(A_{(1)},B_{(1)}) A_{(2)} \cdot B_{(2)}.
\end{align}
The algebra  with respect to  this product is in general no longer Hopf,
but forms a comodule algebra. In general, the branching operator $/\Phi$ is
not an algebra homomorphism and the lack of being homomorphic controls the
new branchings. In \cite{fauser:jarvis:2003a} it was investigated in which
way Hopf algebra cohomology helps to classify branching operators. The
result was roughly, that one  distinguishes 1-cochains which are closed
and  those that are not closed.  The first class yields branching
operators  $/\Phi$  which are homomorphisms, corresponding to  so called
group like series $\Phi$, which are such that
\begin{align}
(s_\lambda\cdot s_\mu)/\Phi = s_\lambda/\Phi \cdot s_\mu/\Phi.
\end{align}
This is related to the fact that in these cases
\begin{align}
\Delta(\Phi) 
&=\sum_{(\Phi)} \Phi_{(1)}\otimes \Phi_{(2)} =\Phi\otimes\Phi
\end{align}
Group like series are of the general form
\begin{align}
\Phi = \prod_{i>0} (1-f(x_i))^{\gamma}
\end{align}
where $f$ is an arbitrary polynomial, and $\gamma$ is in principle
arbitrary (invoking the obvious extension of binomial coefficients in the
complex case). The  epithet `group like' stems from Hopf algebra theory, where
elements with a coproduct $\Delta(g)=g\otimes g$, just doubling the
element, are called group like. The second class of series is more 
interesting. Such series no longer  define homomorphisms  but one can
introduce a 2-cocycle, derived from the 1-cochain, which describes the
deviation from being a homomorphism. In this way Hopf algebra 
cohomology formalises the classification of these branching rules. Further
analysis shows in fact that a cliffordization, that is a Drinfeld twist, underlies this mechanism. Thus 
as mentioned, the structure of symmetric functions of orthogonal and symplectic type is embedded in
deformation theory in the Hopf algebra context. For 
details we refer to \cite{fauser:jarvis:2003a}. 

\subsection{$\pi$-Branchings and products of $\pi$-characters.}
\label{subsec:HopfNewBranchings}

Having introduced the requisite Hopf algebraic structure for the 
understanding of branching rules, we now proceed to revisit the 
combinatorial analysis of proposition \ref{Prop-Mpixy}. Firstly we invoke 
the following properties of plethysms, transcribed into the present 
setting. 
\mybenv{Lemma}\label{Lemma:PlethProperties} 
For any symmetric functions $A$, $B$ and $C$ we have
\begin{align}
     A \otp (B\pm C) =& A\otp B \pm A\otp C, 
     &&& \text{right distributivity,} 
\label{PlethDistributivity} \\
    (A+B) \! \otp \!C =&\, \sum A \! \otp 
    \!C_{(1)} \cdot B \! \otp \!
    C_{(2)}
    &&& \text{left binomial expansion $i)$,}
\label{PlethBinom1} \\
    (A-B) \! \otp \!C =&\, \sum A \! \otp 
    \!C_{(1)} \cdot B \! \otp \!
    {\sf S}(C_{(2)})
    &&& \text{left binomial expansion $ii)$,}
\label{PlethBinom2} \\
    A \otp (B\cdot C) =& (A\otp B)\cdot(A\otp C),
    &&& \text{right homomorphism,} 
\label{RightHom} \\ 
    (A \cdot B) \otp C =&\,\sum A \! \otp \!C_{[1]} \cdot 
    B \! \otp \!C_{[2]},
    &&& \text{left hom. expansion,}
\label{PlethProd} \\
    A \otp (B \otp C) =&\, (A \otp B) \otp C,
    &&& \text{associativity.}
\label{PlethAssoc}
\end{align}
\myeenv
These are standard relations given first by 
Littlewood~\cite{littlewood:1950},
but see also 
\cite{littlewood:1958,littlewood:1958a,littlewood:1958b,macdonald:1979a},
or more recently, \cite{carvalho:dagostino:2001a,carvalho:dagostino:2001b}
as well as \cite{SCHUR},
and the web-site of Brian Wybourne\footnote{%
\texttt{url: http://www.phys.uni.torun.pl/{\~{}bgw/}}}.

As pointed out previously, it is sometimes convenient to use 
an alternative rather suggestive notation for plethysms, by writing
$s_\mu[s_\lambda]=\{\lambda\}\otp\{\mu\}
\equiv \{\lambda\}^{\otp\{\mu\}}$. This latter notation
will be used in our further development, including that of
the following commutativity condition applying to all pairs
of partitions $\lambda$ and $\mu$:
\begin{align}
\Delta(\{\lambda\}^{\otp \{\mu\}}) &=\,
(\Delta(\{\lambda\}))^{\otp \{\mu\}}\,.
 \label{Eq:PlethComm1}
\end{align}
This may be seen by expressing the left hand side of (\ref{Eq:PlethComm1})
in terms of indeterminates $z=(x,y)$, and then noting that 
\begin{align}
  \Delta( \{\lambda\}^{\otp\{\mu\}} )(z)
   = (\{\lambda\}^{\otp\{\mu\}})(x,y)  
   = s_\mu[s_\lambda](x,y) 
   = s_\mu[\Delta(\{\lambda\}](z)
   = ( \Delta(\{\lambda\})(z))^{\otp\{\mu\}}\,,
\end{align}
where this is just the right hand side of (\ref{Eq:PlethComm1}).

Alternatively, setting $x=(x_1,x_2,\ldots,x_m)$ and
$y=(y_1,y_2,\ldots,y_n)$ with $z=(x,y)$ the above outer coproduct of a
plethysm (the left hand side) or plethysm of an outer  coproduct (the right
hand side) may both be interpreted in terms of  branchings from
$\grpGL(m+n)$ to $\grpGL(m)\times \grpGL(n)$.  It is immaterial at what
stage the plethysm evaluation is performed.  In other words there is an
appropriate commuting diagram: 

\begin{align}
\begin{array}{c@{\hskip 2.5truecm}c}
 \rnode{a}{\grpGL(m+n)} 
&\rnode{b}{\grpGL(m+n)}
\\[10ex]
 \rnode{c}{\grpGL(m)\otimes\grpGL(n)}
&\rnode{d}{\grpGL(m)\otimes\grpGL(n)} 
\end{array}
\ncline{->}{a}{b}
\Aput{\otp\{\mu\}}
\ncline{->}{c}{d}
\Aput{\otp\{\mu\}}
\ncline{->}{a}{c}
\Aput{\Delta}
\ncline{->}{b}{d}
\Aput{\Delta}
\end{align}
Or in terms of elements:
\begin{align}
\begin{array}{c@{\hskip 2.5truecm}c}
 \rnode{a}{\{\lambda\}} 
&\rnode{b}{\{\lambda\}\otp\{\mu\}}
\\[10ex]
 \rnode{c}{\Delta(\{\lambda\})}
&\rnode{d}{\Delta(\{\lambda\}\otp\{\mu\})=(\Delta(\{\lambda\}))\otp\{\mu\}} 
\end{array}
\ncline{->}{a}{b}
\Aput{\otp\{\mu\}}
\ncline{->}{c}{d}
\Aput{\otp\{\mu\}}
\ncline{->}{a}{c}
\Aput{\Delta}
\ncline{->}{b}{d}
\Aput{\Delta}
\end{align}

Using this observation we may explitly evaluate the coproduct of a 
plethysm as follows:

\mybenv{Lemma}\label{Lem:coprodPleth} 
The coproduct of the plethysm $\{\lambda\}\otp\{\mu\}$ is  given by
\begin{align}
\Delta(\{\lambda\}\otp \{\mu\}) 
=\sum_{[\mu_{(1)}]}\ldots \sum_{[\mu_{(N^\lambda)}]}
   \sum_{(\mu)^{N^\lambda}} \prod_{k=1}^{N^\lambda}
\left(\{\lambda_{(1)k}\}\otp\{\mu_{(k)[1]}\}\right) \otimes 
\left(\{\lambda_{(2)k}\}\otp\{\mu_{(k)[2]}\}\right) \nn
 \label{Eq:PlethComm2}
\end{align}
where $N^\lambda$ is the number of terms in the outer coproduct of
$\{\lambda\}$. The summations are a variety of Sweedler sums: 
each $\sum_{[\mu]}$ denotes an inner coproduct Sweedler sum, 
while $\sum_{(\mu)^N}$ denotes a multiple outer coproduct Sweedler sum.
\myeenv

\noindent{\bf Proof:}
\begin{align}
\Delta(\{\lambda\}\otp \{\mu\}) &=\,
(\Delta(\{\lambda\}))^{\otp \{\mu\}} 
= \bigg( \sum_{k=1}^{N^\lambda}
\{\lambda_{(1)k}\} \otimes \{\lambda_{(2)k}\}\bigg)^{\otp\{\mu\}}\nn
&= \sum_{(\mu)^{N^\lambda}} \prod_{k=1}^{N^\lambda} 
\left(\{\lambda_{(1)k}\} \otimes \{\lambda_{(2)k}\}\right)^{\otp\{\mu_{(k)}\}}\nn
&= \sum_{[\mu_{(1)}]}\ldots \sum_{[\mu_{(N^\lambda)}]}
   \sum_{(\mu)^{N^\lambda}} \prod_{k=1}^{N^\lambda}
\{\lambda_{(1)k}\}^{\otp\{\mu_{(k)[1]}\}} \otimes 
\{\lambda_{(2)k}\}^{\otp\{\mu_{(k)[2]}\}} \nn
\end{align}
\qed

Generalizations of the above will be required in which multiple
outer coproduct Sweedler sums of the type 
$\sum_{(\Phi)^{N}}$ occur, where $N$ is the number of Sweedler sums
and $\Phi$ is a Schur function series. In the sequel $N$ will often be given,
as here, by the number $N^\lambda$ of terms in the outer coproduct of some
$\lambda$.

We will need the following corollary for the general formula of the
coproduct of the composition of two series. We use `virtual 
representations' \cite{knutson:1973a}, which are necessary to turn the
monoid of representations into an Abelian group \`a la  Grothendieck. The
notion $-\{\mu\}$ is shorthand for $0-\{\mu\}$. A proper definition would
include equivalence classes of pairs of representations under the relation
$\{\mu\}+(-\{\mu\})=0$; this is not displayed here since we need only basic facts
about $-\{\mu\}$.

\mybenv{Corollary}\label{Cor:minusPleth} 
For any symmetric function $C$, partition $\lambda$ and positive  integer
$n$, it follows from the lemma \ref{Lemma:PlethProperties} of plethysms
properties
\begin{align}
0\otp \{\lambda\} &= \delta_{\lambda,(0)}\,\{0\}
\label{eq:0otplambda} \\ 
\nn
0\otp C &= \langle C\mid \{0\} \rangle \,\{0\} 
\label{eq:0otpC} \\
\{0\}\otp \{\lambda\}&= \delta_{\ell(\lambda),1}\ \{0\} =
\bigg\{
\begin{array}{cl}
\{0\} & \text{if~} \ell(\lambda)=1 \text{~that is~}
\{\lambda\}=\{m\}=h_{(m)}   \\
0     & \text{otherwise.} \\
\end{array}
\label{eq:1otplambda} \\
\{0\}\otp C &= \sum_{m\geq0} \langle C\mid h_{(m)} \rangle \, \{0\}
\label{eq:1otpC} \\
(n\{\lambda\})\otp C 
&= (\{\lambda\}+\ldots+\{\lambda\}) \otp C \nn
&= \sum_{(C)^n} \{\lambda\}\otp C_{(1)} \cdot
   \ldots\cdot \{\lambda\}\otp C_{(n)} 
\label{eq:noptC} \\
(-n\{\lambda\})\otp C 
&= (-\{\lambda\}-\ldots-\{\lambda\}) \otp C \nn
&= \sum_{(C)^{n}}
   \{\lambda\}\otp \antip(C_{(1)}) \cdot
   \ldots\cdot \{\lambda\}\otp \antip(C_{(n)}) 
\label{eq:minusnoptC} 
\end{align}
where summation over the Sweedler indices is implicitly assumed, and 
where $\antip$ is the antipode of the outer Hopf algebra.  
\myeenv

In the next theorem we use both outer and  inner coproducts, distinguished
by brackets $(\,\cdot\,)$ and $[\,\cdot\,]$, respectively.

\mybenv{Theorem}\label{The:mainTheorem:i} ({\bf Main Theorem i)}) 
For any Schur function series $\Phi$ and an arbitrary Schur function 
$\{\pi\}$ the outer coproduct of the plethysm $\{\pi\}\otp\Phi$ is
given by:
\begin{align}
\Delta( \{\pi\}\otp \Phi) 
 = \sum_{[\Phi_{(1)}]}\ldots\sum_{[\Phi_{(N^\pi}]}
   \sum_{(\Phi)^{N^\pi}} \prod_{(k)=1}^{N^{\pi}}
       \left( \{\pi_{(1)k}\}\otp \Phi_{(k)[1]} \right) 
\otimes\left( \{\pi_{(2)k}\}\otp \Phi_{(k)[2]} \right) 
\end{align}
\myeenv 

\noindent
{\bf Proof:}
 The result follows immediately from lemma \ref{Lem:coprodPleth} by setting 
$\lambda=\pi$, replacing $\mu$ by $\Phi$ and noting the right
distributivity  of plethysm with respect to addition
(\ref{PlethDistributivity}). 
\qed

Before we draw some conclusions from this theorem, we want to state the
even more general result for the coproduct of the plethysm of two  Schur
function series. Let $\Xi=\sum_\lambda x_\lambda s_\lambda$, with
$x_\lambda\in\openZ$, be a Schur function series.  Such a series is called
Schur positive if all  the coefficients $x_\lambda$ are  nonnegative.  A
Schur function series is called (strictly) Schur negative if all
coefficients are strictly negative. Obviously \emph{any} Schur function series can
be decomposed into a Schur positive and a Schur negative part $\Xi=\Xi^+ -
\Xi^- =\sum_\lambda x^+_\lambda s_\lambda - \sum_\mu x^-_\mu
s_\mu$, with $x_\lambda^+\ge0$ and $x_\mu^->0$. In the following, we assume that there are finitely many nonzero terms, in order to restrict the iterated coproducts to finite depth.
Due to right distributivity of the plethysm the problem can be reduced to
the case $\Xi\otp \{\pi\}$.

\mybenv{Theorem}\label{The:mainTheorem ii} ({\bf Main Theorem ii)})
For any Schur function series $\Xi=\sum_\lambda^{N^+}x^+ s_\lambda - 
\sum_\mu^{N^-}x^- s_\mu$ the outer coproduct of
$\Xi\otp\{\pi\}$ reads
\begin{align}
\Delta( \{\Xi\}\otp \{\pi\})
&=
\sum_{[\pi]} \prod_{k=1}^{N^+} \prod_{l=1}^{x^+_\lambda}
   \{\lambda_{(1)}\} \otp \{\pi_{(1)(k)(l)[1]}\} \otimes
   \{\lambda_{(2)}\} \otp \{\pi_{(1)(k)(l)[2]}\} \nn
&~~~~~~~\cdot  
\sum_{[\pi]} \prod_{k=1}^{N^-} \prod_{l=1}^{x^-_\lambda}
   \{\lambda_{(1)}\} \otp \antip(\{\pi_{(2)(k)(l)[1]}\}) \otimes
   \{\lambda_{(2)}\} \otp \antip(\{\pi_{(2)(k)(l)[2]}\})   
\end{align}
where the outer product of the two tensor factors has been left  unevaluated
for clarity. The composite Sweedler index notation indicates three levels of successive outer coproduct, and a final inner coproduct (see (\ref{eq:MultiSweedler}) below).
\myeenv 

\noindent
{\bf Proof:} 
Using the right distributivity (\ref{PlethDistributivity}) of the
plethysm and corollary \ref{Cor:minusPleth}, the sum splits into
two parts which can be treated by applying lemma \ref{Lem:coprodPleth}. 
\begin{align}
\Delta( \{\Xi\}\otp \{\pi\})
&= \Delta( \{\Xi^+ - \Xi^-\}\otp \{\pi\}) \nn
&= \Delta( \{\Xi^+\}\otp \{\pi_{(1)}\}) 
   \cdot \Delta( \{-\Xi^-\}\otp \{\pi_{(2)}\}) \nn 
&= \Delta( \{\Xi^+\}\otp \{\pi_{(1)}\})
   \cdot \Delta( \{\Xi^-\}\otp \antip(\{\pi_{(2)}\}))
\end{align}
Now we consider the first term
\begin{align}
\Delta(\Xi^+ \otp \pi_{(1)})
&= \Delta(\sum_{\lambda(k),k=1}^{N^+} 
          x^+_\lambda \{\lambda\} \otp \{\pi_{(1)}\} )\nn
&= \Delta(\prod_{k=1}^{N^+} 
          x^+_\lambda \{\lambda\} \otp \{\pi_{(1)(k)}\} )\nn
&= \Delta(\prod_{k=1}^{N^+} \prod_{l=1}^{x^+_\lambda}
   \{\lambda\} \otp \{\pi_{(1)(k)(l)}\} )\nn
&= \sum_{[\pi]} \prod_{k=1}^{N^+} \prod_{l=1}^{x^+_\lambda}
   (\{\lambda_{(1)}\} \otp \{\pi_{(1)(k)(l)[1]}\}) \otimes
   (\{\lambda_{(2)}\} \otp \{\pi_{(1)(k)(l)[2]}\}) \label{eq:MultiSweedler}
\end{align}
and we can expand the second term in an analogous fashion, where the
antipode is given by $\antip(\{\pi\}) = (-1)^{\vert \pi\vert} 
\{\pi^\prime\}$. Hence the result follows.
\qed   

It is clear by now, that the two parts of the Main Theorem allow the
computation of the outer coproduct of the plethysm of a polynomial Schur
function series and another Schur function series
\begin{align}
\Delta(\Xi \otp \Phi)
\end{align}
The resulting formula is of a remarkable complexity and not displayed. 
The interested reader should note, that these coproducts are of polynomial
type, sometimes called Fa\`a di Bruno coproducts. Such coproducts play a
crucial role in renormalization theory and are a typical feature of
coproducts emerging from composition. For references see 
\cite{brouder:frabetti:krattenthaler:2004a}.

\subsection{Application to $M_\pi$}\label{subsec:applicationMPi}

While it would be interesting to examine this general case more  closely,
we want to turn here back to the case of the $M_\pi$ series treated
combinatorially in section \ref{sec:SSeries}. First we need a few more
facts about the $M$ series itself.

\mybenv{Lemma}\label{Lemma:MP} 
The inner coproduct of the $M$ series is given by the Cauchy kernel
(\ref{Eq-Cauchy}):
\begin{align}
\delta M = M_{[1]}\otimes M_{[2]} 
&= \prod_{i,j} (1-x_i y_j)^{-1} 
 = \sum_\sigma s_\sigma(x)s_\sigma(y)
\end{align}
The outer coproduct of the $M$ series is group like:
\begin{align}
\Delta M = M_{(1)}\otimes M_{(2)}
&=\prod_{i} (1-x_i)^{-1} \prod_{j} (1- y_j)^{-1}
 = M\otimes M \\
\Delta^{(k-1)} M &= M_{(1)}\otimes \ldots \otimes M_{(k)} 
                 = M \otimes \ldots \otimes M 
\end{align}
\myeenv

\noindent
{\bf Proof:} These are well known identities, stated or partly proved in 
section \ref{sec:Heuristics}, see (\ref{Eq-Cauchy}) and
(\ref{Eq-Mxy}). The final formula can either be seen directly from the 
product form of $M$ or deduced recursively.
\qed

\mybenv{Corollary}\label{Corr:Mpi} 
[Specialization of Main Theorem i), see proposition \ref{Prop-Mpixy}]
\newline
For any partition $\pi$, the coproduct of 
$M_\pi=\{\pi\}\otp M \equiv \{\pi\}^{\otp M}$ 
is given by:

\begin{align}
\Delta M_\pi
&= (M_\pi)_{(1)}\otimes (M_\pi)_{(2)} \label{Eq-Mpi12}\\
&= (M_\pi\otimes M_\pi) \cdot 
  (\{\pi'_{(1)}\}^{\otp M_{[1]}}
 \otimes\{\pi'_{(2)}\}^{\otp M_{[2]}})  \label{Eq-Mpi12cut}\\ 
&= (M_\pi\otimes M_\pi) \cdot
\prod_{\xi,\eta<\pi}\prod_{k=1}^{C^\pi_{\xi\eta}}\sum_{\sigma(\xi,\eta,k)} 
(\{\xi\}\otp\{\sigma(\xi,\eta,k)\}) \otimes
(\{\eta\}\otp\{\sigma(\xi,\eta,k\}) \label{Eq-Mpicoprod}
\end{align}
where the $C^\pi_{\xi\eta}$ are the Littlewood-Richardson coefficients
of outer multiplication. The summations over the $\sigma(\xi,\eta,k)$ 
are formally over all Schur functions.
\myeenv

\noindent
{\bf Proof:}
We give an independent proof, since this is the main technical
result required to obtain the $M_\pi$ branchings. First,
it should be noted that (\ref{Eq-Mpi12}) is just Sweedler
notation for the decomposition of the coproduct of $M_\pi$.
Its evaluation serves to define the Sweedler sum
$(M_\pi)_{(1)}\otimes (M_\pi)_{(2)}$. Using the suggestive
exponential notation, the second form (\ref{Eq-Mpi12cut})
comes about by noting that, since $M_{(k)}=M$ for all $k$,
we have
\begin{align}
\Delta M_\pi &= \Delta(\{\pi\}^{\otp M}) =(\Delta(\{\pi\}))^{\otp M} \nn 
&= (\{\pi\}\otimes\{0\}+\{0\}\otimes\{\pi\}+\Delta'(\{\pi\}))^{\otp M}
\nn
&=(\{\pi\}\otimes\{0\})^{\otp M} \cdot
  (\{0\}\otimes\{\pi\})^{\otp M} \cdot
  \Delta'(\{\pi\}))^{\otp M} 
\label{Eq-MpiDprime}\end{align}
where 'proper cuts' (\ref{PK}) have been used.
The second factor of (\ref{Eq-Mpi12cut}) is just
$\Delta'(\{\pi\}))^{\otp M}$ expressed in terms of Sweedler
sums. To obtain the first factor, it should be noted that 
as a consequence of (\ref{RightHom}), lemma~\ref{Lemma:MP} 
and (\ref{eq:1otplambda}), we have
\begin{align}
\label{Eq:pi0M}
(\{\pi\}\otimes\{0\})^{\otp M} 
= \sum_{\sigma} \{\pi\}^{\otp\{\sigma\}} \otimes
\{0\}^{\otp\{\sigma\}} 
= \sum_{m} \{\pi\}^{\otp \{m\}}\otimes\{0\} 
= \{\pi\}^{\otp M}\otimes\{0\}=M_\pi\otimes\{0\}.
\end{align}
Similarly $(\{0\}\otimes\{\pi\})^{\otp M}=\{0\}\otimes M_\pi$.
Hence
\begin{align}
(\{\pi\}\otimes\{0\})^{\otp M} \cdot (\{0\}\otimes\{\pi\})^{\otp M} 
=  M_\pi\otimes\{0\} \cdot \{0\}\otimes M_\pi=M_\pi\otimes M_\pi,
\end{align}
as required in (\ref{Eq-MpiDprime}) to give the first factor
of (\ref{Eq-Mpi12cut}). It only remains to calculate the
cut coproduct plethysm. This is given by 
\begin{align}
 \Delta'(\{\pi\}))^{\otp M} 
&= \bigl(\sum_{\xi,\eta<\pi} C^\pi_{\xi\eta} 
  \{\xi\}\otimes\{\eta\}\bigr)^{\otp M} 
= \prod_{\xi,\eta<\pi}\prod_{k=1}^{C^\pi_{\xi\eta}}\ 
    \bigl(\{\xi\}\otimes\{\eta\}\bigr)^{\otp M} \nn
&= \prod_{\xi,\eta<\pi}\prod_{k=1}^{C^\pi_{\xi\eta}} 
  \sum_{\sigma(\xi,\eta,k)}
    \{\xi\}^{\otp\sigma(\xi,\eta,k)} \otimes
    \{\eta\}^{\otp\sigma(\xi,\eta,k)},
\end{align}
precisely as required\footnote{In (\ref{eq:CutInverses}) below this expression is denoted $M_{\pi^\prime_{(1)}} \otimes M_{\pi^\prime_{(2)}}$.}.
\qed

Reverting to the Schur function notation for plethysms and introducing
indeterminates $z=(x,y)$ in connection with  the coproducts, it is clear
that (\ref{Eq-Mpicoprod}) coincides with (\ref{Eq-Mpixy}).  Examples of
such coproducts have already been displayed in example  \ref{Ex:DMpi}.

The skewing by a series amounts to establishing a branching rule from
$\grpGL(n)$ characters, that is Schur functions, into $\grpH_\pi$ 
characters of a subgroup $\grpH_\pi(n)\subset \grpGL(n)$ which leaves a
tensor of symmetry type $\pi$ invariant.  Such branching rules take the
form (\ref{Eq-brHpi}), that is
$\{\lambda\}\rightarrow\(\lambda/M_\pi\)_\pi$. 
At this point the relation between the structure of the Schur function series
and the nature of the branching, mentioned in the introductory remarks to this section,
can be discerned. While group like series
induce branching operators which are algebra  homomorphisms, those series
$M_\pi$ with nontrivial plethysms  $\vert\pi\vert\ge 2$, are in general no
longer homomorphisms (see section \ref{subsec:Kernels} and appendix \ref{sec:GraphCalc} for further details).
The following theorem allows a generalization
to arbitrary series  $M_\pi$.

\mybenv{Theorem}
For any partitions $\mu$, $\nu$ and series $M_\pi$ one finds the
$\pi$-skewed branching formula
\begin{align}
(\{\mu\}\cdot\{\nu\})/M_\pi
&= \{\mu/(M_\pi)_{(1)}\}\cdot \{\nu/(M_\pi)_{(2)}\} \nn
&= \sum_{\sigma(\xi,\eta,k)}
\{\mu/(M_\pi \prod_{\xi\eta<\pi}\prod_{k=1}^{C^\pi_{\xi\eta}}
        \xi\otp \sigma(\xi,\eta,k)\}
\cdot
\{\nu/(M_\pi \prod_{\xi\eta<\pi}\prod_{k=1}^{C^\pi_{\xi\eta}}
        \eta\otp \sigma(\xi,\eta,k)\}
\end{align}  
where the summations over the $\sigma(\xi,\eta,k)$ are formally over all Schur
functions, and, as usual, $C^\pi_{\xi,\eta}$ indicates the 
number of terms $\{\xi\}\otimes\{\eta\}$ appearing in the 
coproduct of $\{\pi\}$ for fixed $\xi$ and $\eta$.
\myeenv

\noindent
{\bf Proof:}
We make use of the duality mediated by the Schur scalar product
\begin{align}
  \langle( s_\mu\cdot s_\nu)/M_\pi\mid s_\rho\rangle
&= \langle s_\mu\cdot s_\nu\mid M_\pi \cdot s_\rho\rangle \nn
&= \langle s_\mu\otimes s_\nu\mid \Delta(M_\pi \cdot s_\rho)\rangle \nn
&= \langle s_\mu\otimes s_\nu\mid \Delta(M_\pi)\cdot\Delta(s_\rho)\rangle \nn
&= \langle s_\mu\otimes s_\nu\mid
           ( (M_\pi)_{(1)}\otimes (M_\pi)_{(2)} )\cdot\Delta(s_\rho)\rangle \nn
&= \langle s_\mu/(M_\pi)_{(1)}\otimes s_\nu/(M_\pi)_{(2)}\mid
   \Delta(s_\rho)\rangle \nn
&= \langle s_\mu/(M_\pi)_{(1)}\cdot s_\nu/(M_\pi)_{(2)}\mid  
   s_\rho\rangle
\end{align}
Since the Schur scalar product is non degenerate, and since $s_\rho$ was
arbitrary and the Schur functions form a basis the result 
follows, with the final step made by identifying
(\ref{Eq-Mpi12}) with (\ref{Eq-Mpicoprod}). 
This proof does not make use of the property that $M_\pi$
is group like, and thus applies quite generally for all $\pi$.\qed

Before we can prove the $\pi$-generalized version of the 
Newell-Littlewood theorem we need a Lemma about the coproduct
of inverse $M_\pi$ series.
\mybenv{Lemma} \label{l1}
Every $M_\pi$ series has as inverse the $L_\pi$ series, furthermore, 
the part obtained by proper cuts of the coproduct is invertible too
\begin{align}
M_\pi\cdot L_\pi &= 1 \label{eq:Inverses} \\
M_{\pi^\prime_{(1)}}L_{\pi^\prime_{(1)}}\otimes 
M_{\pi^\prime_{(2)}}L_{\pi^\prime_{(2)}} 
&= 1\otimes 1 \label{eq:CutInverses}
\end{align}
\myeenv

\noindent
{\bf Proof:} From the observation that $M\cdot L=1$, we obtain
\begin{align}
M_\pi\cdot L_\pi &=
      (\{\pi\}\otp M)\cdot (\{\pi\}\otp L) 
\,=\, \{\pi\}\otp (M\cdot L)
\,=\, \{\pi\} \otp \{0\} 
\,=\, \{0\}
\,=\, 1
\end{align}
An alternative way to see this reads
\begin{align}
M_\pi\cdot L_\pi 
&= \{\pi\}\otp M \cdot (-\{\pi\})\otp M
 = \sum_{(M)} \{\pi\}\otp M_{(1)} \cdot (-\{\pi\})\otp M_{(2)} \nn
&= (\{\pi\}-\{\pi\})\otp M = 0\otp M = \delta_{\{0\},M} = 1
\end{align}
showing that using the identities of the algebra of plethysm, 
equations (\ref{PlethBinom1}) and  (\ref{eq:0otpC}), is much weaker
than the first consideration, since we had to make use of 
the fact that $M$ is group like. In particular this shows, that 
the plethysm by a negative Schur function $-\{\pi\}$ applied to
a series $\Phi$ gives the inverse $(\Phi_\pi)^{-1}$ if and only if 
$\Phi$ is group like. The previous result can be used to 
establish the following identity
\begin{align}
1\otimes1 \,=\, \{0\}\otimes \{0\} &= \Delta(\{0\})
\,=\, \Delta(M_\pi\cdot L_\pi) 
\,=\, \Delta(M_\pi)\cdot \Delta(L_\pi)\nn
&=    (M_{\pi_{(1)}}\otimes M_{\pi_{(2)}})\cdot
      (L_{\pi_{(1)}}\otimes L_{\pi_{(2)}}) \nn
&=    (M_{\pi_{(1)}}\cdot L_{\pi_{(1)}}) \otimes 
      (M_{\pi_{(2)}}\cdot L_{\pi_{(2)}})  \nn
&=
(M_{\pi}M_{\pi^\prime_{(1)}}L_{\pi}L_{\pi^\prime_{(1)}})\otimes
(M_{\pi}M_{\pi^\prime_{(2)}}L_{\pi}L_{\pi^\prime_{(2)}}) \nn
&= 
(M_{\pi}L_{\pi}M_{\pi^\prime_{(1)}}L_{\pi^\prime_{(1)}})\otimes
(M_{\pi}L_{\pi}M_{\pi^\prime_{(2)}}L_{\pi^\prime_{(2)}}) \nn
&=
(M_{\pi^\prime_{(1)}}L_{\pi^\prime_{(1)}})\otimes
(M_{\pi^\prime_{(2)}}L_{\pi^\prime_{(2)}})
\end{align}
where one should remember that primes at Sweedler indices denote 
the proper cuts of the coproduct.\qed

Products of characters of $\grpH_\pi(n)\subset \grpGL(n)$ can be obtained
using the following generalization of the Newell-Littlewood theorem.

\mybenv{Theorem}\label{The-munuHn}
[$\pi$-Newell-Littlewood theorem, proposition \ref{Prop-pHpi}]
Let $\dbrace{\mu}{\pi}$, $\dbrace{\nu}{\pi}$ be formal universal characters
of $\grpH_{\pi}(n)$ defined in terms of Schur functions by 
$\(\mu\)_\pi=\{\mu/L_\pi\}$ and $\(\nu\)_\pi=\{\nu/L_\pi\}$, as in 
(\ref{Eq-brinvHpi}). 
Then 
\begin{align}
\dbrace{\mu}{\pi}\cdot\dbrace{\nu}{\pi} 
&= \( \mu/ \bigl(\{\pi'_{(1)}\}^{\otp M_{[1]}}\bigr)
   \cdot
      \nu/ \bigl(\{\pi'_{(2)}\}^{\otp M_{[2]}}\bigr) \)_\pi \nn 
&= \sum_{\sigma(\xi,\eta,k)}
\dbrace{\{\mu/(\prod_{\xi\eta<\pi}\prod_{k=1}^{C^\pi_{\xi\eta}}
          \{\xi\}\otp \sigma(\xi,\eta,k)
\}\cdot
   \{\nu/(\prod_{\xi\eta<\pi}\prod_{k=1}^{C^\pi_{\xi\eta}}
          \{\eta\}\otp \sigma(\xi,\eta,k)
\}
}{\pi}
\label{Eq-munuHn}
\end{align}
where the $\sigma(\xi,\eta,k)$ that are associated with the Sweedler
indices are formally summed over all Schur functions.
\myeenv
\noindent
{\bf Proof:} We make use of duality, and of the second identity of lemma 
\ref{l1} to calculate the product of  $\grpH_\pi(n)$ characters directly in terms of Schur functions:
\begin{align}
\langle \dbrace{\mu}{\pi}\cdot \dbrace{\nu}{\pi} \mid s_\rho \rangle
&=
\langle \mu\otimes \nu \mid L_\pi \otimes L_\pi \cdot \Delta\,s_\rho \rangle \nn
&=\langle \mu\otimes \nu \mid
(M_{\pi^\prime_{(1)}}L_{\pi^\prime_{(1)}})\otimes
(M_{\pi^\prime_{(2)}}L_{\pi^\prime_{(2)}})
L_\pi\otimes L_\pi \cdot \Delta\,s_\rho \rangle \nn
&=\langle \mu\otimes \nu \mid
M_{\pi^\prime_{(1)}}\otimes M_{\pi^\prime_{(2)}} \,\cdot\,
L_{\pi^\prime_{(1)}}\otimes L_{\pi^\prime_{(2)}} \,\cdot\,
L_\pi\otimes L_\pi\cdot\Delta\,s_\rho \rangle \nn
&=
\langle \mu/M_{\pi^\prime_{(1)}}\otimes \nu/M_{\pi^\prime_{(2)}} \mid
\Delta L_\pi \cdot \Delta s_\rho \rangle \nn
&=
\langle \mu/M_{\pi^\prime_{(1)}}\otimes \nu/LM_{\pi^\prime_{(2)}} \mid
\Delta (L_\pi \cdot s_\rho) \rangle \nn
&=
\langle 
\dbrace{\mu/M_{\pi^\prime_{(1)}}\,\cdot\,\nu/M_{\pi^\prime_{(2)}}}{\pi}\mid
s_\rho \rangle
\end{align}
The conclusion follows from nondegeneracy of the Schur scalar product, and completeness of the Schur 
basis.
\qed

\noindent
{\bf Proof${\mathbf '}$:} Alternatively, and more directly, by virtue of the $\grpGL(n)\supset \grpH_\pi(n)$ branching rule (\ref{Eq-brHpi}) and its inverse (\ref{Eq-brinvHpi}), the multiplicity
of the $\grpH_\pi(n)$ character $\(\rho\)_\pi$ in the product
$\(\mu\)_\pi\cdot\(\nu\)_\pi$ is given in terms of characters of
$\grpGL(n)$, that is to say Schur functions, by:
\begin{align}
\langle \dbrace{\mu}{\pi}\cdot \dbrace{\nu}{\pi} \mid
 \dbrace{\rho }{\pi}\rangle &= \langle\{ ( \mu/L_\pi \cdot \nu/L_\pi )/M_\pi\} \mid {\{}\rho/L_\pi {\}} \rangle \nn
&= \langle\{ \mu/L_\pi \cdot \nu/L_\pi \} \mid M_\pi \cdot {\{}\rho/L_\pi {\}} \rangle \nn
&= \langle \{\mu/L_\pi\}\otimes\{\nu/L_\pi\} \mid
  \Delta(M_\pi \cdot {\{}\rho/L_\pi {\}}) \rangle \nn
&= \langle \{\mu/L_\pi\}\otimes\{\nu/L_\pi\} \mid
  \Delta(M_\pi)\cdot \Delta({\{}\rho/L_\pi {\}}) \rangle \nn
&= \langle \{\mu/L_\pi\}\otimes\{\nu/L_\pi\} \mid
  (M_\pi\otimes M_\pi)\cdot (\{\pi'_{(1)}\}^{\otp M_{[1]}}
 \otimes\{\pi'_{(2)}\}^{\otp M_{[2]}})\cdot
 \Delta({\{}\rho/L_\pi {\}}) \rangle \nn
&= \langle \{\mu/ (L_\pi M_\pi (\{\pi'_{(1)}\}^{\otp M_{[1]}} )) \}
\otimes
   \{\nu/L_\pi M_\pi (\{\pi'_{(2)}\}^{\otp M_{[2]}} )) \}
\mid \Delta({\{}\rho/L_\pi {\}}) \rangle \nn
&= \langle \{\mu/ (\{\pi'_{(1)}\}^{\otp M_{[1]}}) \}
\cdot
   \{\nu/ (\{\pi'_{(2)}\}^{\otp M_{[2]}}) \}
\mid {\{}\rho/L_\pi {\}} \rangle\end{align}
Here (as in the first proof above) use has been made in particular of the 
proper cut coproduct plethysm (\ref{Eq-Mpi12cut}), evaluated by means  of
(\ref{Eq-Mpicoprod}).  However note that although the inverse series
(equation (\ref{eq:Inverses})) are required, the cut product inverse
series are not invoked (equation (\ref{eq:CutInverses})). Again the
nondegeneracy of  the Schur scalar product and the completeness of the Schur 
basis, allow us  immediately to draw the conclusion (\ref{Eq-munuHn}), which
is in accord with proposition~\ref{Prop-pHpi}. This proof is a direct
application of definition (\ref{prodPhi}) for $\Phi = M_\pi$ and $\Phi^{-1} 
= L_\pi$.
\qed

\subsection{\label{subsec:Kernels}Scalar products, Cauchy kernels
and plethystic generalisations}

In the present section we discuss briefly the role of the Cauchy kernel
(\ref{Eq-Cauchy}) and its inverse, the Cauchy-Binet formula and their 
generalizations in the branching process. In appendix \ref{sec:GraphCalc}
we give an additional description in terms of tangles which we consider
to be helpful in contemplating this structure.

The coproducts of group like branchings induce algebra homomorphisms, which
are nontrivial in the sense that they still induce new representations, 
see the $V$ series in table \ref{eq:series}, but lie in the trivial cohomology
class  in the sense 
of algebra deformation theory.
\begin{align}
(\Delta\Phi)(x,y) 
&= \Phi(x)\Phi(y) \label{eq:grpcoprd} \\
m_\phi(x\otimes y) 
&= (\partial\phi)(x_{(1)},y_{(1)}) x_{(2)}\cdot y_{(2)} \nn
&= \Phi^{-1}(\Phi(x)\cdot\Phi(y))
\end{align}
The relation between the Schur function series and the linear forms used
in branching operators (\ref{brOp}) is given via the Schur scalar product
$\phi = (\Phi \vert . )$. However, if $\grpO(n)$ or $\grpSp(n)$ are 
considered, one is confronted with branchings which are mediated by branching
operators which fail to be group like. From the discussion above, we know how
the coproduct of series is computed and that the Cauchy kernel plays a key role
in the derivation of these results. Moreover, due to plethysms, convolution
products of the Cauchy kernel appear, where in one slot or both slots a 
plethysm takes place.

While the Cauchy kernel (\ref{Eq-Cauchy}) presents a $0\rightarrow 2$ map,
producing from the 'zero series' $1$ a second rank tensor, the Schur scalar
product is just a $2\rightarrow 0$ map which takes a second rank tensor to a
number.
\begin{align}
1 &\rightarrow \prod_{i,j} (1-x_i y_j)^{-1} 
   = \sum_\sigma u_\sigma\otimes v_\sigma \\
   x\otimes y &\rightarrow (x\mid y)
\end{align}
Note that $u_\sigma$, $v_\sigma$ is a dual pair of bases with respect to the Schur 
scalar product. Hence the scalar product and Cauchy kernel are dual objects,
this is graphically displayed in equation (\ref{A:eval}). This opens the 
possibility of studying scalar products which are dual to the coevaluations
used in the coproduct formulae of Schur function series. Such a view amounts
to introducing a deformation in the product, not in the coproduct. Moreover,
this duality explains why the Cauchy kernel
\begin{align}
{\sf C}(x,y) &:= \prod_{i,j} (1-x_i y_j)^{-1} 
   = \sum_\sigma u_\sigma\otimes v_\sigma 
\end{align}
is a reproducing kernel
for Schur function (series) -- for, if $F$ is an arbitrary symmetric function, we have   
\begin{align}
{\sf C} \cdot F (x)   
        &:= \sum_\sigma u_\sigma(x)  \langle v_\sigma\mid F \rangle
   \equiv  F(x). 
\end{align}
Displayed in the graphical notation this property resembles a Reidemeister 
move\footnote{In fact, the variable $y$ can be considered here to be a linear form
on $x$ and the action written $ \mbox{`}\sum_{(y)} {\sf C}(x,y) F(y)\mbox{'}$.}. 

Before we discuss scalar products we consider coproducts with primitive
elements, that is power sum symmetric functions. The outer coproduct of a
one part power symmetric function $p_n$ has exactly two parts
\begin{align}
\Delta(p_n) &= p_n\otimes 1 + 1 \otimes p_n
\end{align}
From our Main Theorem i) (Theorem \ref{The:mainTheorem:i}) we obtain the
fact that any power sum plethysm by a group like series $G$ 
remains group like,
\begin{align}
\Delta( \{p_n\}^{\otp G}) 
&={\Delta(p_n)}^{\otp G} = ({p_n\otimes p_0+ p_0 \otimes p_n})^{\otp G}  \nn
&= (p_n\otimes p_0)^{\otp G}  \cdot (p_0 \otimes p_n)^{\otp G} \nn
& = \{p_n\}^{\otp G} \otimes \{p_n\}^{\otp G},
\end{align}  
where in the last line, only the first Sweedler term in the inner coproduct $\delta(\{\lambda\}) = \{\lambda\} \otimes \{|\! \lambda\! | \} + \cdots $
survives. However, due to the inner coproduct this fails for plethysms
of the form $\{p_{\lambda}\}\otp G$ for $\lambda$ a partition having more than
one part.

The interesting terms which occur for non group like coproducts induce a
twisting exactly such that the lack of being a homomorphism is compensated. This
can be seen explicitly in proposition \ref{Prop-Mpixy}, where the right hand side has two
terms $M_\pi(x)M_\pi(y)$ resembling the part which would be group like, and the
twist $\prod\prod\sum s_{\xi\otp \sigma}(x) s_{\eta\otp\sigma}(y)$, related to
the proper cuts of $\Delta(\pi)$. Due to the general theory of such 
deformations and the fact that we are dealing with group characters, we can 
conclude that this twisting is a 2-cycle, its dual is also a 2-cocycle. 
Furthermore, we note that the $M$ series is the unit of the inner product, and
that the Cauchy kernel can be obtained by applying the inner coproduct to the 
$M$ series, $\delta M(x,y)={\sf C}(x,y)$. This is the way the Cauchy kernel 
entered in the branching formulae derived above.

Not all plethysms give rise to different 2-cycles. For example, since  the proper cut
part of the coproduct of $\{2\}$ and $\{1^2\}$ are identical, they induce the
same 2-cycle, that is the Cauchy kernel. This shows, that there are homologous
2-cycles, and a major question is to classify all such cycles.

Switching to the dual setting, we consider $2\rightarrow 0$ maps, which
generalize the Schur scalar product. For example, for the plethystic deformation 
using $\{1^3\}$, we obtain a new scalar product
\begin{align}
(x\mid y)_{\{1^3\}} 
&= \langle x_{(1)} \mid \{1^2\}\otp y_{(1)} \rangle 
   \langle \{1^2\}\otp x_{(2)} \mid y_{(2)} \rangle
\end{align}
while the Schur scalar product is the special case
\begin{align}
\langle x\mid y \rangle
&= (x\mid y)_{\{2\}}
 = (x\mid y)_{\{1^2\}}
\end{align}
The trivial case $(x\mid y)_{\{1\}}=\delta_{x,\{0\}}\delta_{y,\{0\}}$ 
belongs to the group like case. Dualizing the twist for a general $\pi$ 
introduced in proposition \ref{Prop-Mpixy} we obtain the family of scalar
products
\begin{align}
(x\mid y)_\pi
&= \prod_{\xi,\eta}\prod^{C^\pi}
   \langle \{\xi\}\otp x\mid \{\eta\}\otp y \rangle
\end{align}
which in general is easily seen to be nonisomorphic and noncohomologous to the
Schur scalar product. All these scalar products can be used to introduce twists
and all of these scalar products induce associative product deformations, since
we are dealing with group characters, hence are 2-cocycles.

Our brief discussion shows that the new branchings open also a new research
area in deformation theory, since we have a hand on the mechanism to introduce
families of appropriate 2-cocycles. A major problem is to classify these
2-cocycles, and to provide tools to be able to decide if two such 2-cocycles are
cohomologous. Moreover, the identification of the groups related to these
2-cocycles is important, in relation to the question of whether these families exhaust the space of
cohomology classes of 2-cocycles. Some further discussion has been put into
appendix \ref{sec:GraphCalc}.

\section{Conclusions}
\setcounter{equation}{0}

The present work opens the door into a fascinating new field of group
representation techniques. Starting with some Hopf algebraically motivated 
questions posed in \cite{fauser:jarvis:2003a} it was possible to derive
group branchings tied to the plethystic series of $M_\pi$ and $L_\pi$
type. Branchings based on these series led by direct calculations to 
$\grpSL$ groups considered as subgroups of appropriate $\grpGL$ groups,
$\grpSL(n)\equiv \grpH_{\{1^n\}}\subset\grpGL(n)$. While the general theory
works with formal characters in the inductive limit, finite representation
spaces require modification rules to cope with syzygies. This is beyond the
scope of the present work, but hints were given in the finite dimensional
examples. These examples uncovered non semi simple and non reductive groups, such as affine groups, and more general semi-direct product groups. A case
study was presented for $\grpH_{1^3}(3)\equiv\grpSL(3)$ and 
$\grpH_{1^3}(4)\supset\grpGL(3)\times\grpGL(1)$. Character tables,
modification rules and products of characters were derived.

We paused the combinatorial exploration to introduce the Hopf algebra
machinery which proved to be a powerful way of in encoding the complexity of the
new branchings. To our knowledge, these branchings have not appeared in the
literature before, since a purely combinatorial route to them would be rather
difficult to discern. However, using Hopf algebraic branching operators, we can tie
to any Schur function series a branching process. If we restrict ourselves
to $M_\pi$ series, these are matrix subgroups of $\grpGL(n)$ which fix a
tensor of Young symmetry $T^\pi$. Particular cases are the branchings
$\grpGL(n)\downarrow\grpGL(n-1)$, $\grpGL(n)\downarrow\grpO(n)$ and
$\grpGL(2n)\downarrow\grpSp(2n)$ fixing a vector $v_i$, a symmetric tensor
$g_{ij}=g_{ji}$, and an antisymmetric tensor $f_{ij}=-f_{ji}$. The $\grpSL$
groups appear through the same mechanism as $\grpH_{1^n}(n)\subset \grpGL(n)$
fixing a volume form (antisymmetric highest rank tensor in $n$ dimension).

The Main Theorem parts i) and ii) (Theorems \ref{The:mainTheorem:i} and 
\ref{The:mainTheorem ii}) allow in principle the choice of a linear combination
of tensors of different Young symmetry type. The formulae obtained have a
close relation to vertex operators, which are composed from two Schur
function series $\Gamma(z)=H(z)E^\perp(z^{-1})$, see
\cite{macdonald:1979a}. This  supports our finding that affine groups can
occur. The question about canonical forms of tensors of a certain Young
type arises, and of their  physical relevance.

Related to this is the question of whether Hall-Littlewood symmetric functions can be 
related to the plethystic scalar products, which we defined in section
\ref{subsec:Kernels}. Starting from a nontrivial group like branching, one
obtains scalar products which have a kernel establishing a residual symmetry
in the Schur function series, see the $V$ series in table (\ref{eq:series}).
It is known that the freedom of choosing $q$ to be a root of unity is related
to representation theory over finite geometries. This observation should be
contrasted with the arbitrary introduction of $q$-generalisations via a braided grade group, which has
no \emph{a priori} geometrical meaning.
The Hopf algebraic treatment may also open a way to understand the
modification rules of the involved groups in a systematic way, so that specific
case can be examined in the light of more general results.

Finally, the Hopf algebraic treatment does not make explicit use of the ground field.
Hence our methods are in principle applicable to $G$-sets, working in the
Burnside ring, and representations of pro-finite groups on them. This leads to
the realm of modular representation theory, and should provide deep insights and
beautiful combinatorics.

From a physical point of view, we want to emphasize that groups fixing
higher rank tensors are related to non-linear models. For example the
antisymmetric tensor  $\epsilon_{abcd}$ in $n$ dimensions is saturated by
$4$ `fields' $\psi^a$. The related invariants are no longer of binary type,
and hence necessarily non linear. Such tensors provide `interaction terms'
like  $\epsilon_{abcd}\psi^b\psi^c\psi^d$ in the field equations. There is thus a
host of possible applications for our methods, which we  hope to explore
in future work.

\noindent{\bf Acknowledgment:} \\ 
PDJ and BF acknowledge the Australian Research Council, research grant
DP0208808, for partial support. They also thank the Alexander von Humboldt 
Foundation for a `sur place' travel grant to BF for a visit to the 
University of Tasmania, where part of this work was done, and also the School of Mathematics and Physics for hospitality.
RCK is pleased to acknowledge the award of a Leverhulme Emeritus Fellowship
supporting in part this collaboration.
\begin{appendix}

\section{\label{sec:GraphCalc}Graphical calculus for plethystic (co)scalar
products}
\setcounter{equation}{0}

In this appendix we provide enough notion of graphical calculus 
\cite{yetter:1990a,lyubashenko:1995a,lyubashenko:1995b,kuperberg:1991a,kuperberg:1996a}
to be able to discuss the structure of the 2-cocycles and coscalar products
involved in the branching process of the plethystic branchings. The method consist
in a graphical representation of the index structure of tensors. A detailed exposition
may be found in \cite{fauser:2002c}.

Let $W$ be a (finite dimensional) complex vector space and $W^*$ its dual.
Choose two (nonintersecting) horizontal lines, an input line and an output
line, where two sets of a nonnegative number of vertices are positioned.
We depict $W$ by a line with downward orientation\footnote{%
This is the ``pessimistic arrow of time'', as coined by Z. Oziewicz in his
talk at the ICCA6, 1999 in Ixtapa.} 
connecting an upper and lower vertex point, and depict $W^*$ by a similar line with 
upward orientation.
\begin{align}
\pspicture[0.5](0,0)(1,2.5)
\psset{linewidth=\pstlw,xunit=0.5,yunit=0.5,runit=0.5}
\psset{arrowsize=2pt 2,arrowinset=0.2}
\psline{-}(1,5)(1,0)
\psline{->}(1,4)(1,2.5)
\rput(0,2.75){$W$}
\endpspicture 
\hskip 1truecm
\pspicture[0.5](0,0)(1,2.5)
\psset{linewidth=\pstlw,xunit=0.5,yunit=0.5,runit=0.5}
\psset{arrowsize=2pt 2,arrowinset=0.2}
\psline{-}(1,5)(1,0)
\psline{->}(1,1)(1,2.5)
\rput(0,2.75){$W^*$}
\endpspicture
\hskip 1truecm
\pspicture[0.5](0,0)(1,2.5)
\psset{linewidth=\pstlw,xunit=0.5,yunit=0.5,runit=0.5}
\psset{arrowsize=2pt 2,arrowinset=0.2}
\psline{-}(1,5)(1,0)
\psline{->}(1,5)(1,3.5)
\pscircle[linewidth=0.4pt,fillstyle=solid,fillcolor=black](1,2.5){0.2}
\psline{->}(1,2.5)(1,1.5)
\rput(0,2.75){$S$}
\endpspicture
\hskip 1truecm
\pspicture[0.5](0,0)(4,2.5)
\psset{linewidth=\pstlw,xunit=0.5,yunit=0.5,runit=0.5}
\psset{arrowsize=2pt 2,arrowinset=0.2}
\psline{-}(1,5)(1,3.5)
\psline{->}(1,5)(1,4)
\psline{-}(2,5)(2,3.5)
\psline{->}(2,5)(2,4)
\psline[linestyle=dotted]{-}(3,4.5)(6,4.5)
\psline{-}(7,5)(7,3.5)
\psline{->}(7,5)(7,4)
\psline{-}(0.5,3.5)(7.5,3.5)
\psline{-}(0.5,3.5)(0.5,1.5)
\psline{-}(0.5,1.5)(7.5,1.5)
\psline{-}(7.5,1.5)(7.5,3.5)
\psline{-}(1,0)(1,1.5)
\psline{->}(1,1)(1,0.5)
\psline{-}(2,0)(2,1.5)
\psline{->}(2,1)(2,0.5)
\psline[linestyle=dotted]{-}(3,0.5)(6,0.5)
\psline{-}(7,0)(7,1.5)
\psline{->}(7,1)(7,0.5)
\rput(5,2.5){$T$}
\endpspicture
\hskip 1truecm
\pspicture[0.5](0,0)(1.5,2.5)
\psset{linewidth=\pstlw,xunit=0.5,yunit=0.5,runit=0.5}
\psset{arrowsize=2pt 2,arrowinset=0.2}
\psline{-}(1,5)(1,3.5)
\psline{->}(1,5)(1,4)
\psline{-}(2,5)(2,3.5)
\psline{->}(2,3.5)(2,4.5)
\psline{-}(0.5,3.5)(2.5,3.5)
\psline{-}(0.5,3.5)(0.5,1.5)
\psline{-}(0.5,1.5)(2.5,1.5)
\psline{-}(2.5,1.5)(2.5,3.5)
\psline{-}(1,0)(1,1.5)
\psline{->}(1,0)(1,1)
\psline{-}(2,0)(2,1.5)
\psline{->}(2,1)(2,0.5)
\rput(1.5,2.5){$U$}
\endpspicture
\end{align}
\begin{align}
\Id_{W}
\hskip 1truecm
\Id_{W^*}
\hskip 1truecm
S^{\mu}_{\nu}
\hskip 1.5truecm
T^{\mu_1\ldots \mu_k}_{\nu_1\ldots \nu_l}
\hskip 2.5truecm
U^{\nu_1\mu_2}_{\mu_1\nu_2}
\end{align}
and all summations of these tensors are over the $\nu$ indices (inputs).
Operations of maps are represented as dots or boxes in these 
tangles\footnote{%
One may read these diagrams as a type of flow chart, if the mathematical 
background of monoidal tensor categories is neglected.}. The number of input
and output lines of a map may differ, one speaks then of an $k\rightarrow l$
map. Products and coproducts are special such maps of valence $2\rightarrow 1$
and $1\rightarrow 2$
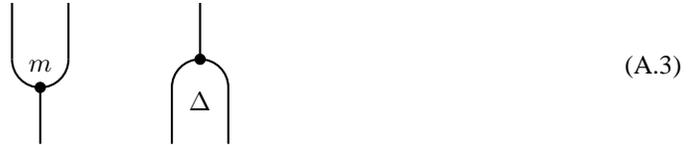
\begin{align}
\pspicture[0.5](0,0)(1.5,2.5)
\psset{linewidth=\pstlw,xunit=0.5,yunit=0.5,runit=0.5}
\psset{arrowsize=2pt 2,arrowinset=0.2}
\psline{-}(0.5,5)(0.5,3)
\psline{-}(2.5,5)(2.5,3)
\psarc(1.5,3){1.0}{180}{360}
\pscircle[linewidth=0.4pt,fillstyle=solid,fillcolor=black](1.5,2){0.2}
\psline{-}(1.5,2)(1.5,0)
\rput(1.5,2.75){$m$}
\endpspicture
\hskip 1truecm
\pspicture[0.5](0,0)(1.5,2.5)
\psset{linewidth=\pstlw,xunit=0.5,yunit=0.5,runit=0.5}
\psset{arrowsize=2pt 2,arrowinset=0.2}
\psline{-}(1.5,5)(1.5,3)
\psarc(1.5,2){1.0}{0}{180}
\pscircle[linewidth=0.4pt,fillstyle=solid,fillcolor=black](1.5,3){0.2}
\psline{-}(0.5,2)(0.5,0)
\psline{-}(2.5,2)(2.5,0)
\rput(1.5,1.5){$\Delta$}
\endpspicture
\end{align} 
where we have omitted the orientation. In the case of symmetric functions, 
the product is the outer product of symmetric functions and the coproduct 
is the outer coproduct of symmetric functions. The space $W=\oplus 
V^{\otimes^n}$ is the graded space of symmetric functions in infinitely
many variables and $V$ is the grade one space generated by the variables
$x_i$.

We use the \textit{same graphical notation} for series of Schur functions
too. Hence we depict the $M$ series or an $A,B,C,D,\ldots,X,\ldots$ series
also by a single line. This can be done due to the fundamental theorem of
symmetric functions \cite{knutson:1973a}, which allows us to regard bases,
such as  the elementary symmetric functions, complete symmetric functions, 
Schur functions etc, as new generators of the graded space $W$. The product
of two $M$ series is hence given as
\begin{align}
m( M(x) \otimes M(y)) 
&\cong M(x,y)\vert_{x=y} \,=\, M(x,x)\nn
\sum_m \{m\} \cdot \sum_n \{n\} 
&= \sum_{n,m} \{m\}\cdot \{n\}
 = \sum_{n,m} \sum_\pi C^\pi_{m,n} \{\pi\}
\end{align}
which is a new series but may not have a standard name like $A,B,C,\ldots$.
The same holds true for coproducts. Before we consider such coproducts, we
introduce a graphical symbol for the plethysm operation on a Schur function
$W[s_\pi]\equiv \{\pi\}\otp W$. Furthermore we need the tangles for
evaluation, coevaluation, scalar products and coscalar products, \emph{i.e.} the
tangle for the Cauchy kernel $u_i\otimes v_i$ where $u_i$ and $v_i$ are
mutually dual bases $\langle u_i\mid v_j \rangle = \delta_{i,j}$. We depict all this as
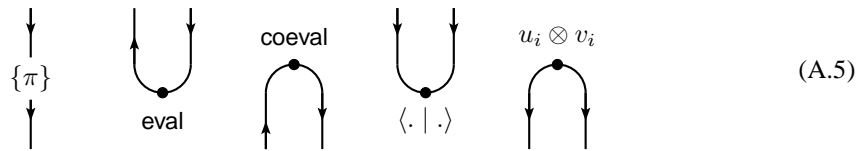
\begin{align}\label{A:eval}
\pspicture[0.5](0,0)(1,2.5)
\psset{linewidth=\pstlw,xunit=0.5,yunit=0.5,runit=0.5}
\psset{arrowsize=2pt 2,arrowinset=0.2}
\psline{-}(1,5)(1,0)
\psline{->}(1,5)(1,4)
\psline{->}(1,2)(1,1)
\rput(1,2.5){\psframebox*[framesep=3pt]{$\{\pi\}$}}
\endpspicture 
\hskip 1truecm
\pspicture[0.5](0,0)(1,2.5)
\psset{linewidth=\pstlw,xunit=0.5,yunit=0.5,runit=0.5}
\psset{arrowsize=2pt 2,arrowinset=0.2}
\psline{-}(0,5)(0,3)
\psline{->}(0,3)(0,4)
\psline{-}(2,5)(2,3)
\psline{->}(2,5)(2,4)
\psarc(1,3){1.0}{180}{360}
\pscircle[linewidth=0.4pt,fillstyle=solid,fillcolor=black](1,2){0.2}
\rput(1,1){$\textsf{\small eval}$}
\endpspicture
\hskip 1truecm
\pspicture[0.5](0,0)(1,2.5)
\psset{linewidth=\pstlw,xunit=0.5,yunit=0.5,runit=0.5}
\psset{arrowsize=2pt 2,arrowinset=0.2}
\psline{-}(0,0)(0,2)
\psline{->}(0,0)(0,1)
\psline{-}(2,0)(2,2)
\psline{->}(2,2)(2,1)
\psarc(1,2){1.0}{0}{180}
\pscircle[linewidth=0.4pt,fillstyle=solid,fillcolor=black](1,3){0.2}
\rput(1,4){$\textsf{\small coeval}$}
\endpspicture
\hskip 1truecm
\pspicture[0.5](0,0)(1,2.5)
\psset{linewidth=\pstlw,xunit=0.5,yunit=0.5,runit=0.5}
\psset{arrowsize=2pt 2,arrowinset=0.2}
\psline{-}(0,5)(0,3)
\psline{->}(0,5)(0,4)
\psline{-}(2,5)(2,3)
\psline{->}(2,5)(2,4)
\psarc(1,3){1.0}{180}{360}
\pscircle[linewidth=0.4pt,fillstyle=solid,fillcolor=black](1,2){0.2}
\rput(1,1){$\langle.\mid .\rangle$}
\endpspicture
\hskip 1truecm
\pspicture[0.5](0,0)(1,2.5)
\psset{linewidth=\pstlw,xunit=0.5,yunit=0.5,runit=0.5}
\psset{arrowsize=2pt 2,arrowinset=0.2}
\psline{-}(0,0)(0,2)
\psline{->}(0,2)(0,1)
\psline{-}(2,0)(2,2)
\psline{->}(2,2)(2,1)
\psarc(1,2){1.0}{0}{180}
\pscircle[linewidth=0.4pt,fillstyle=solid,fillcolor=black](1,3){0.2}
\rput(1,4){$u_i\otimes v_i$}
\endpspicture
\end{align}
Now we start to depict the coproduct for the $M$ series (or any group
like series $Y\in \{L,M,V,\ldots\}$) and that of a series $X\in 
\{A,B,C,D,\ldots\}$.
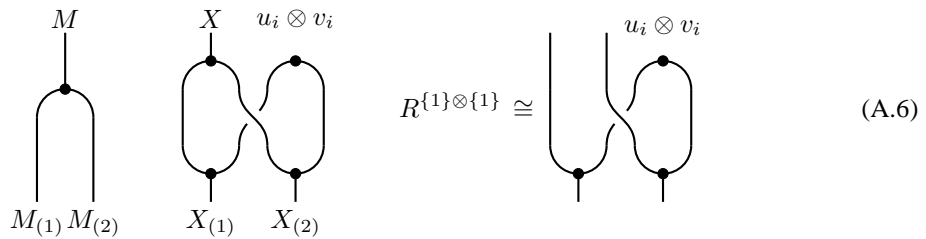
\begin{align}\label{RMatrixX}
\pspicture[0.5](0,0)(1.5,4)
\psset{linewidth=\pstlw,xunit=0.5,yunit=0.5,runit=0.5}
\psset{arrowsize=2pt 2,arrowinset=0.2}
\psline{-}(1.5,7)(1.5,5)
\psarc(1.5,4){1.0}{0}{180}
\pscircle[linewidth=0.4pt,fillstyle=solid,fillcolor=black](1.5,5){0.2}
\psline{-}(2.5,4)(2.5,1)
\psline{-}(0.5,4)(0.5,1)
\rput(1.5,7.5){$M$}
\rput(0.5,0.25){$M_{(1)}$}
\rput(2.5,0.25){$M_{(2)}$}
\endpspicture
\hskip 1truecm
\pspicture[0.5](0,0)(2.5,4)
\psset{linewidth=\pstlw,xunit=0.5,yunit=0.5,runit=0.5}
\psset{arrowsize=2pt 2,arrowinset=0.2}
\psline{-}(1,7)(1,6)
\psarc(1,5){1.0}{0}{180}
\pscircle[linewidth=0.4pt,fillstyle=solid,fillcolor=black](1,6){0.2}
\psarc(4,5){1.0}{0}{180}
\pscircle[linewidth=0.4pt,fillstyle=solid,fillcolor=black](4,6){0.2}
\psline{-}(0,5)(0,3)
\psline{-}(5,5)(5,3)
\psbezier(3,5)(3,4)(2,4)(2,3)
\psbezier[border=3pt,bordercolor=white](2,5)(2,4)(3,4)(3,3)
\psarc(1,3){1.0}{180}{360}
\pscircle[linewidth=0.4pt,fillstyle=solid,fillcolor=black](1,2){0.2}
\psarc(4,3){1.0}{180}{360}
\pscircle[linewidth=0.4pt,fillstyle=solid,fillcolor=black](4,2){0.2}
\psline{-}(1,2)(1,1)
\psline{-}(4,2)(4,1)
\rput(1,7.5){$X$}
\rput(4,7.5){$u_i\otimes v_i$}
\rput(1,0.25){$X_{(1)}$}
\rput(4,0.25){$X_{(2)}$}
\endpspicture
\hskip 1truecm
R^{\{1\}\otimes\{1\}}\,\cong\,\,
\pspicture[0.5](0,0)(2.5,4)
\psset{linewidth=\pstlw,xunit=0.5,yunit=0.5,runit=0.5}
\psset{arrowsize=2pt 2,arrowinset=0.2}
\psline{-}(0,7)(0,5)
\psline{-}(2,7)(2,5)
\psarc(4,5){1.0}{0}{180}
\pscircle[linewidth=0.4pt,fillstyle=solid,fillcolor=black](4,6){0.2}
\psline{-}(0,5)(0,3)
\psline{-}(5,5)(5,3)
\psbezier(3,5)(3,4)(2,4)(2,3)
\psbezier[border=3pt,bordercolor=white](2,5)(2,4)(3,4)(3,3)
\psarc(1,3){1.0}{180}{360}
\pscircle[linewidth=0.4pt,fillstyle=solid,fillcolor=black](1,2){0.2}
\psarc(4,3){1.0}{180}{360}
\pscircle[linewidth=0.4pt,fillstyle=solid,fillcolor=black](4,2){0.2}
\psline{-}(1,2)(1,1)
\psline{-}(4,2)(4,1)
\rput(4,7.25){$u_i\otimes v_i$}
\endpspicture
\end{align}
This shows that the deformation for $X$ like series comes up with an
$R$-matrix which is built from the Cauchy kernel. Hence the coproduct
of $X$ like series can be understood as induced by an $R$-matrix
\begin{align}
\Delta_X 
&= R^{\{1\}\otimes\{1\}}\circ \Delta,
  \qquad \forall X\in \{A,B,C,D,\ldots\}
\end{align}
The elements $X$ are all given by plethysms of the form $A=\{1^2\}\otp M$,
$B=\{1^2\}\otp L$, $C=\{2\}\otp L$, $D=\{2\}\otp M$, in general by a plethysm 
with $\{1^2\}$ or $\{2\}$ of a group like series $Y$. That these two plethysms
act in the same way stems from the fact that they possess the same proper cut 
coproduct 
\begin{align}
\Delta^\prime \{2\} 
&= \Delta^\prime \{1^2\}
 = \{1\} \otimes \{1\} 
\end{align}
The $R$-matrix related to these coproducts is given by the plethysms of
a sufficient number of Cauchy kernels (here one) with the proper cut coproduct
of the involved Schur function $\{\pi\}$. This reads as tangle
\begin{align}\label{RMatrix}
 R^{\Delta^\prime\{2\}}
=R^{\Delta^\prime\{1^2\}}
=R^{\{1\}\otimes\{1\}}
&\cong
 \,\,
\pspicture[0.5](0,0)(2.5,4)
\psset{linewidth=\pstlw,xunit=0.5,yunit=0.5,runit=0.5}
\psset{arrowsize=2pt 2,arrowinset=0.2}
\psline{-}(0,7)(0,5)
\psline{-}(2,7)(2,5)
\psarc(4,5){1.0}{0}{180}
\pscircle[linewidth=0.4pt,fillstyle=solid,fillcolor=black](4,6){0.2}
\psline{-}(0,5)(0,3)
\psline{-}(5,5)(5,3)
\psbezier(3,5)(3,4)(2,4)(2,3)
\psbezier[border=3pt,bordercolor=white](2,5)(2,4)(3,4)(3,3)
\psarc(1,3){1.0}{180}{360}
\pscircle[linewidth=0.4pt,fillstyle=solid,fillcolor=black](1,2){0.2}
\psarc(4,3){1.0}{180}{360}
\pscircle[linewidth=0.4pt,fillstyle=solid,fillcolor=black](4,2){0.2}
\psline{-}(1,2)(1,1)
\psline{-}(4,2)(4,1)
\rput(4,7.25){$u_i\otimes v_i$}
\rput(3,5){\psframebox*{$\scriptscriptstyle \{1\}$}}
\rput(5,5){\psframebox*{$\scriptscriptstyle \{1\}$}}
\endpspicture
\end{align}
Since the deformed coproducts are derived from group characters we know, that 
these coproducts are coassociative. This allows one to conclude that the involved 
2-chains are actually 2-cycles. The trivial 2-cycle $\eta\otimes\eta$ yields 
a trivial deformation $R=1\otimes 1$. The next nontrivial case is given by
$R^{\{1\}\otimes\{1\}}= u_i\otimes v_i$, \emph{i.e.} induced by the Cauchy kernel 
itself. The new branchings induced by plethysms of weight greater than 2 are 
then obtained by applying the proper cut coproduct elements as plethysms to
a sufficient number of coevaluations (Cauchy kernels). The resulting formula
is that of proposition \ref{Prop-Mpixy}, resp. corollary \ref{Corr:Mpi}.
If the proper cut coproduct terms of $\pi$ are indexed by $a\ldots xy$,
this deformation reads as a tangle
\begin{align}\label{weepingwillow}
\pspicture[0.5](0,0)(6,5.5)
\psset{linewidth=\pstlw,xunit=0.5,yunit=0.5,runit=0.5}
\psset{arrowsize=2pt 2,arrowinset=0.2}
\psline{-}(1,10)(1,9)
\psarc(1,8){1.0}{0}{180}
\psline{-}(0,8)(0,3)
\psline{-}(2,8)(2,5)
\psline{-}(3,8)(3,5)
\psline{-}(11,8)(11,6)
\psarc(4,8){1.0}{0}{180}
\psarc(7,8){1.0}{0}{180}
\psarc(10,8){1.0}{0}{180}
\psarc(6,6){1.0}{180}{270}
\psarc(10,6){1.0}{180}{360}
\psarc(8,5){1.0}{180}{270}
\psarc(1,3){1.0}{180}{270}
\psbezier(5,6)(5,7)(6,7)(6,8)
\psbezier(6,5)(7,5)(9,7)(9,8)
\psbezier(8,4)(9,4)(10,4)(10,5)
\psbezier(1,2)(2,2)(4,3)(4,4)
\psbezier(5,2)(6,2)(8,3)(8,4)
\psarc(4,5){1.0}{180}{270}
\psbezier(4,4)(5,4)(6,4)(6,5)
\psline{-}(1,2)(1,1)
\psline{-}(5,2)(5,1)
\psbezier[border=3pt,bordercolor=white](2,5)(2,2)(4,2)(5,2)
\psbezier[border=3pt,bordercolor=white](5,8)(5,7)(7,6)(7,5)
\psbezier[border=3pt,bordercolor=white](8,8)(8,7)(9,7)(9,6)
\psline[linestyle=dotted]{-}(5,9)(6,9)
\psline[linestyle=dotted]{-}(4.5,4.25)(5.5,4.75)
\psline[linestyle=dotted]{-}(8.5,4.25)(9.5,4.75)
\pscircle[linewidth=0.4pt,fillstyle=solid,fillcolor=black](1,9){0.2}
\pscircle[linewidth=0.4pt,fillstyle=solid,fillcolor=black](4,9){0.2}
\pscircle[linewidth=0.4pt,fillstyle=solid,fillcolor=black](7,9){0.2}
\pscircle[linewidth=0.4pt,fillstyle=solid,fillcolor=black](10,9){0.2}
\pscircle[linewidth=0.4pt,fillstyle=solid,fillcolor=black](6,5){0.2}
\pscircle[linewidth=0.4pt,fillstyle=solid,fillcolor=black](10,5){0.2}
\pscircle[linewidth=0.4pt,fillstyle=solid,fillcolor=black](4,4){0.2}
\pscircle[linewidth=0.4pt,fillstyle=solid,fillcolor=black](8,4){0.2}
\pscircle[linewidth=0.4pt,fillstyle=solid,fillcolor=black](1,2){0.2}
\pscircle[linewidth=0.4pt,fillstyle=solid,fillcolor=black](5,2){0.2}
\pspolygon[linewidth=0.4pt,linestyle=dashed](2.5,9.5)(2.5,3.5)(11.5,3.5)(11.5,9.5)
\rput(3,8){\psframebox*[border=2pt]{$\scriptscriptstyle \pi_{1a}$}}
\rput(4.75,8){\psframebox*[border=2pt]{$\scriptscriptstyle \pi_{2a}$}}
\rput(6,8){\psframebox*[border=2pt]{$\scriptscriptstyle \pi_{1x}$}}
\rput(7.75,8){\psframebox*[border=2pt]{$\scriptscriptstyle \pi_{2x}$}}
\rput(9,8){\psframebox*[border=2pt]{$\scriptscriptstyle \pi_{1y}$}}
\rput(11,8){\psframebox*[border=2pt]{$\scriptscriptstyle \pi_{2y}$}}
\rput(1,10){\psframebox*[border=2pt]{$\scriptscriptstyle M_{\pi}$}}
\rput(1,0.25){\psframebox*[border=2pt]{$\scriptscriptstyle M_{\pi(1)}$}}
\rput(5,0.25){\psframebox*[border=2pt]{$\scriptscriptstyle M_{\pi(2)}$}}
\endpspicture
\hskip 1truecm
\Leftrightarrow
\hskip 1truecm
\pspicture[0.5](0,0)(2.5,4)
\psset{linewidth=\pstlw,xunit=0.5,yunit=0.5,runit=0.5}
\psset{arrowsize=2pt 2,arrowinset=0.2}
\psline{-}(1,7)(1,6)
\psarc(1,5){1.0}{0}{180}
\pscircle[linewidth=0.4pt,fillstyle=solid,fillcolor=black](1,6){0.2}
\psarc(4,5){1.0}{0}{180}
\pscircle[linewidth=0.4pt,fillstyle=solid,fillcolor=black](4,6){0.2}
\psline{-}(0,5)(0,3)
\psline{-}(5,5)(5,3)
\psbezier(3,5)(3,4)(2,4)(2,3)
\psbezier[border=3pt,bordercolor=white](2,5)(2,4)(3,4)(3,3)
\psarc(1,3){1.0}{180}{360}
\pscircle[linewidth=0.4pt,fillstyle=solid,fillcolor=black](1,2){0.2}
\psarc(4,3){1.0}{180}{360}
\pscircle[linewidth=0.4pt,fillstyle=solid,fillcolor=black](4,2){0.2}
\psline{-}(1,2)(1,1)
\psline{-}(4,2)(4,1)
\rput(1,7.5){$M_\pi$}
\rput(4,7.5){$u^\pi_i\otimes v^\pi_i$}
\rput(1,0.25){$M_{\pi(1)}$}
\rput(4,0.25){$M_{\pi(2)}$}
\endpspicture
\end{align}
which might be called ``weeping willow'' diagram. The framed part of this 
tangle is again a $0\rightarrow 2$ tangle, hence a 2-cycle or coscalar 
product denoted as $u^\pi_i\otimes v^\pi_i$. These 2-cycles are in general
homologically different, however, there are degeneracies as 
$(u)^{\Delta^\prime\{2\}}= u^{\{1\}}_i\otimes v^{\{1\}}_i =
 (u)^{\Delta^\prime\{1^2\}}$ clearly shows. Of course, all of the above
arguments dualize to yield a theory of 2-cocycles and deformed products.
A major question which has to be solved is to classify these scalar
products. 

\section{\label{sec:combProofs}Combinatorial proofs of propositions
\ref{Prop-skewMpi} and \ref{Prop-pHpi}}
\setcounter{equation}{0}

To prove in a combinatorial way proposition \ref{Prop-skewMpi} we require, 
in addition to our previous combinatorial development, hence using no
Hopf formalism, a pair of lemmas.
\mybenv{Lemma}
\label{Lem-opskew}
For any partitions $\lambda$, $\mu$ and $\nu$
\begin{align}
   (\{\mu\}\,\{\nu\})\,/\,\{\lambda\}
&=\sum_{\sigma,\tau}\ C^\lambda_{\sigma\tau}\ \{\mu/\sigma\}\ \{\nu/\tau\}. 
\label{Eq-opskew}
\end{align}
\myeenv

\noindent{\bf Proof:}
In order to prove this we must be more precise about
what the left-hand side really means. It can be written 
more explicitly in the form
\begin{align}
 (\{\mu\}\,\{\nu\})\,/\,\{\lambda\}
&=\left(s_\mu(x)\,s_\nu(x)\right)\,/\, s_\lambda(x) 
 =\sum_{\rho}\ C^\rho_{\mu\nu}\ s_{\rho/\lambda}(x).
\label{Eq-rho}
\end{align}
Now consider the following two expansions of the same Schur
function products:
\begin{align}
  s_\mu(x,y)\ s_\nu(x,y) 
&= \sum_{\rho}\ C^\rho_{\mu\nu}\ s_\rho(x,y) 
\ =\ \sum_{\rho,\lambda} C^\rho_{\mu\nu}\ s_{\lambda/\rho}(x)\ s_\lambda(y)\cr
&= \sum_\lambda s_\lambda(y) 
      \left( \sum_{\rho} C^\rho_{\mu\nu}\ s_{\lambda/\rho}(x)  \right).
\label{Eq-exp1} 
\end{align}
and
\begin{align}
  s_\mu(x,y)\ s_\nu(x,y) 
&= \sum_{\sigma,\tau}\ s_{\mu/\sigma}(x)\ s_\sigma(y)\
     s_{\nu/\tau}(x)\ s_\tau(y)\cr
&= \sum_{\sigma,\tau}\ C^\lambda_{\sigma\tau}\ 
     s_{\mu/\sigma}(x)\ s_{\nu/\tau}(x)\ s_\lambda(y)\cr
&= \sum_\lambda s_\lambda(y)
     \left(  \sum_{\sigma,\tau}\ C^\lambda_{\sigma\tau}\ 
     s_{\mu/\sigma}(x)\ s_{\nu/\tau}(x) \right).
\label{Eq-exp2}
\end{align}
Comparing these two expansions we see that
\begin{align}
\sum_{\rho} C^\rho_{\mu\nu}\ s_{\lambda/\rho}(x)
&= \sum_{\sigma,\tau}\ C^\lambda_{\sigma\tau}\ 
     s_{\mu/\sigma}(x)\ s_{\nu/\tau}(x).
\label{Eq-compare}
\end{align}
Thanks to (\ref{Eq-rho}) this is all that is required to
prove Lemma~\ref{Lem-opskew}.
\qed

Our second lemma is a linear extension of this:
\mybenv{Lemma}
\label{Lem-opZskew}
Let $Z(x)$ be any series of Schur functions
\begin{align}
   Z(x) &=\sum_{\lambda}\ z_\lambda\ s_\lambda(x),
\label{Eq-Z}
\end{align}
and let the coproduct of $Z$ take the form
\begin{align}
   Z(x,y) &=\sum_{\sigma,\tau}\ C^Z_{\sigma\tau}\ s_\sigma(x)\ s_\tau(y).
\label{Eq-Zxy}
\end{align}
Then
\begin{align}
(\{\mu\}\,\{\nu\})\,/\, Z 
&= \sum_{\sigma,\tau}\ C^Z_{\sigma\tau}\ \{\mu/\sigma\}\ \{\nu/\tau\}.
\label{Eq-opZ}
\end{align}
\myeenv

\noindent{\bf Proof:}
First note that
\begin{align}
   Z(x,y)&=\sum_{\lambda}\ z_\lambda\ s_\lambda(x,y)
    = \sum_{\lambda,\sigma,\tau}\ z_\lambda\ C^\lambda_{\sigma\tau}\ 
      s_\sigma(x)\ s_\tau(y), 
\end{align}
so that
\begin{align}
   C^Z_{\sigma\tau} &=\sum_{\lambda}\ z_\lambda\ C^\lambda_{\sigma\tau}.
\label{Eq-cZ}
\end{align}
Then, using Lemma~\ref{Lem-opskew}
\begin{align}
   (\{\mu\}\,\{\nu\})\,/\, Z 
   &= \sum_{\lambda}\ z_\lambda\ (s_\mu(x)\,s_\nu(x))\,/\,s_\lambda(x)\cr
   &= \sum_{\lambda}\ z_\lambda\ \sum_{\sigma\tau}\ 
     C^\lambda_{\sigma\tau}\ s_{\mu/\sigma}(x)\ s_{\nu/\tau}(x)\cr 
   &= \sum_{\sigma,\tau}\ C^Z_{\sigma\tau}\ \{\mu/\sigma\}\ \{\nu/\tau\}.
\label{Eq-opZsk}
\end{align}
as required.
\qed

{\bf Proof of proposition \ref{Prop-skewMpi}:}
Returning to proposition~\ref{Prop-skewMpi} its proof amounts 
to replace $Z$ by $M_\pi$ in lemma~\ref{Lem-opZskew}, and 
substituting into it the $M_\pi(x,y)$ decomposition from 
proposition~\ref{Prop-Mpixy}. This gives
\begin{align}
M_\pi(x,y)
&=
\sum_{\sigma,\tau}\ C^{M_\pi}_{\sigma\tau}\ s_\sigma(x)\ s_\tau(y)\cr
&=
M_\pi(x)\,M_\pi(y)\ 
  \prod_{0\neq\xi,\eta\neq\pi}\ 
  \prod_{k=1}^{C^\pi_{\xi\eta}}\
  \sum_{\sigma(\xi,\eta,k)}\
  s_{\xi\otimes\sigma(\xi,\eta,k)}(x)\ s_{\eta\otimes\sigma(\xi,\sigma,k)}(y).
\label{Eq-MpiZ}
\end{align}
Using this in (\ref{Eq-opZ}) with $Z=M_\pi$ then gives
(\ref{Eq-skewMpi}) as required to prove proposition~\ref{Prop-skewMpi}.
\qed 

\noindent{\bf Proof of proposition \ref{Prop-pHpi}:}
To evaluate the required product of formal characters of $\grpH_\pi(n)$,
one first expresses them in terms of characters of $\grpGL(n)$ using
(\ref{Eq-brinvHpi}), carries out the product in $\grpGL(n)$
and then expresses the resulting Schur functions back in terms
of formal characters of $\grpH_\pi(n)$ using the branching 
rule~(\ref{Eq-brHpi}). This procedure gives
\begin{align}
\hskip-1cm
\(\mu\)\,\(\nu\)\ 
&=
\{\mu/M_\pi^{-1}\}\,\{\nu/M_\pi^{-1}\}
\ =\ \( (\{\mu/M_\pi^{-1}\}\,\{\nu/M_\pi^{-1}\})/M_\pi\)\\
&= 
\sum_{\sigma(\xi,\eta,k)}\ 
  \( \{\mu/(M_\pi^{-1}\,M_\pi\,
  \prod_{{0\neq\xi}\atop{\eta\neq\pi}} 
  \prod_{k=1}^{C^\pi_{\xi\eta}}
  \xi\otp\sigma(\xi,\eta,k)\} \nn
&\qquad\qquad\qquad\cdot\quad
  \{\nu/(M_\pi^{-1}\,M_\pi\,
  \prod_{{0\neq\xi}\atop{\eta\neq\pi}} 
  \prod_{k=1}^{C^\pi_{\xi\eta}}
  \eta\otp\sigma(\xi,\eta,k)\} \)\\
&= 
\sum_{\sigma(\xi,\eta,k)}\ 
  \( \{\mu/
  \prod_{{0\neq\xi}\atop{\eta\neq\pi}}   
  \prod_{k=1}^{C^\pi_{\xi\eta}}
  \xi\otp\sigma(\xi,\eta,k)\}\ 
  \{\nu/
  \prod_{{0\neq\xi}\atop{\eta\neq\pi}} 
  \prod_{k=1}^{C^\pi_{\xi\eta}}
  \eta\otp\sigma(\xi,\eta,k)\} \),
\label{Eq-pHpi-proof}
\end{align}
which completes the proof.\qed

\section{\label{Tables}Tables}
\setcounter{equation}{0}

\subsection{$\grpH_{3}(4)$ formal characters}

\paragraph{Dimensions of $\grpGL(4)\downarrow\grpH_{3}(4)$ irreps:}
\begin{align}
\begin{array}{|l|l|l|l|}
\hline
\{\lambda\}_{\textrm{dim}} & \(\lambda/M_3\)_{\textrm{dim}} & 
\{\lambda\}_{\textrm{dim}} & \(\lambda/M_3\)_{\textrm{dim}} \\ \hline
\{0\}_1 	&  \(0\)_1 	&
\{1\}_{4} 	&  \(1\)_4 \\
\{11\}_6 	&  \(11\)_6 	&
\{111\}_{4} 	&  \(111\)_4 \\
\{1^4\}_1 	&  \(1^4\)_1 	&
\{1^5\}_0 	&  \(1^5\)_0 \\
\{1^6\}_0 	&  \(1^6\)_0 	&
\{2\}_{10} 	&  \(2\)_{10} \\
\{21\}_{20} 	&  \(21\)_{20} 	&
\{211\}_{15} 	&  \(211\)_{15} \\
\{2111\}_{4} 	&  \(2111\)_{4} 	&
\{21^4\}_{0} 	&  \(21^4\)_{0} \\
\{22\}_{20} 	&  \(22\)_{20} 	&
\{221\}_{20} 	&  \(221\)_{20} \\
\{2211\}_{6} 	&  \(2211\)_{6} 	&
\{22111\}_{0} 	&  \(22111\)_{0} \\
\{3\}_{20} 	&  \(3\)_{19}+\(0\)_{1} 	&
\{31\}_{45} 	&  \(31\)_{41}+\(1\)_{4} \\
\{311\}_{36} 	&  \(311\)_{30}+\(11\)_{6} 	&
\{3111\}_{10} 	&  \(3111\)_{6}+\(111\)_{4} \\
\{31^4\}_{0} 	&  \(31^4\)_{-1}+\(1^4\)_{1} 	&
\{32\}_{60} 	&  \(32\)_{50}+\(2\)_{10} \\
\{321\}_{64} 	&  \(321\)_{44}+\(21\)_{20} 	&
\{3211\}_{20} 	&  \(3211\)_{5}+\(211\)_{15} \\
\{32111\}_{0} 	&  \(32111\)_{-4}+\(2111\)_{4} 	&
\{33\}_{50} 	&  \(33\)_{31}+\(3\)_{19} \\
\{331\}_{60} 	&  \(331\)_{19}+\(31\)_{41} 	&
\{3311\}_{20} 	&  \(3311\)_{-10}+\(311\)_{30} \\
\{33111\}_{0} 	&  \(33111\)_{-6}+\(3111\)_{6} 	&
\{331^4\}_{0} 	&  \(331^4\)_{1}+\(31^4\)_{-1} \\
\{4\}_{35} 	&  \(4\)_{31}+\(1\)_{4} 	&
\{41\}_{84} 	&  \(41\)_{68}+\(2\)_{10}+\(11\)_{6} \\
\{411\}_{70} 	&  \(411\)_{46}+\(21\)_{20}+\(111\)_{4} 	&
\{4111\}_{20} 	&  \(4111\)_{4}+\(211\)_{15}+\(1^4\)_{1} \\
\{41^4\}_{0} 	&  \(41^4\)_{-4}+\(2111\)_{4}+\(1^5\)_0 	&
\{42\}_{126}	&  \(42\)_{86}+\(3\)_{19}+\(21\)_{20}+\(0\)_1 \\ 
\hline
\end{array}
\end{align}

From this table one could start to derive modification rules. If we denote a
fully symmetric tensor as $\eta$, then $\(\lambda_1,\lambda_2,\ldots,
\lambda_n\)$, some $\lambda_i\ge3$ can be contracted with $\eta$. Hence the
branching $\{3\}_{20}\downarrow \(3\)_{19}+\eta\(0\)_{1}$ extracts a  triply
contracted 'trace' with respect to $\eta$. However, there is still some
freedom, since we find 20 rank three fully symmetric tensors in  dimension 4.
A more detailed investigation of these affairs is postponed for another
publication, since without a theoretical device this task is tied to tedious
case by case studies.

\paragraph{Some examples of product formulae for $\grpH_{3}(4)$ characters:}

\begin{align}
\begin{array}{c|l}
\cdot & \(1\)_{4} \\
\hline\hline
\(1\)_{4}  & \(2\)_{10}+\(11\)_{6} \\
\(2\)_{10} & \(3\)_{19}+\(21\)_{20}+{\red\(0\)_{1}} \\ 
\(11\)_{6} & \(21\)_{20}+\(111\)_{4} \\
\(3\)_{19} & \(4\)_{31}+\(31\)_{41}+{\red\(1\)_{4}} \\
\(21\)_{20}& \(31\)_{41}+\(22\)_{20}+\(211\)_{15}+{\red\(1\)_{4}} \\
\(111\)_{4}& \(21^2\)_{15}+\(1^4\)_{1} \\ 
\end{array}
\nonumber
\end{align}

\begin{align}
\begin{array}{c|l}
\cdot & \(2\)_{10} \\
\hline\hline
\(2\)_{10}  & \(4\)_{31}+\(31\)_{41}+\(22\)_{20}+{\red2\(1\)_{4}} \\ 
\(11\)_{6}  & \(31\)_{41}+\(211\)_{15}+{\red\(1\)_{4}} \\
\(3\)_{19}  & \(5\)_{46}+\(41\)_{68}+\(32\)_{50}+{\red2\(2\)_{10}+\(11\)_{6}} \\
\(21\)_{20} & \(41\)_{68}+\(32\)_{50}+\(311\)_{30}+\(221\)_{20}
       +{\red2\(2\)_{10}+2\(11\)_{6}} \\
\(111\)_{4} & \(311\)_{30}+\(2111\)_{4}
       +{\red\(11\)_{6}} \\ 
\end{array}
\nonumber
\end{align}

\begin{align}
\begin{array}{c|l}
\cdot & \(11\)_{6} \\
\hline\hline
\(11\)_{6} & \(22\)_{20}+\(211\)_{15}+\(1^4\)_{1} \\
\(3\)_{19} & \(41\)_{68}+\(311\)_{30}+{\red\(2\)_{10}+\(11\)_{6}} \\
\(21\)_{20}& \(32\)_{50}+\(311\)_{30}+\(221\)_{20}+\(2111\)_{4}
      +{\red\(2\)_{10}+\(11\)_{6}} \\
\(111\)_{4}& \(221\)_{20}+\(2111\)_{4} \\ 
\end{array}
\nonumber
\end{align}

\begin{align}
\begin{array}{c|l}
\cdot & \(3\)_{19} \\
\hline\hline
\(3\)_{19}  & \(6\)_{64}+\(51\)_{101}+\(42\)_{86}+\(33\)_{31}
       +{\red2\(3\)_{19}+2\(21\)_{20}+\(0\)_{1}} \\ 
\(21\)_{20} & \(51\)_{101}+\(42\)_{86}+\(411\)_{46}+\(321\)_{44}
       +{\red2\(3\)_{19}+3\(21\)_{20}+\(111\)_{4}+\(0\)_{1}} \\
\(111\)_{4} & \(411\)_{46}+\(3111\)_{6}
        +{\red\(21\)_{20}+\(111\)_{4}} \\ 
\end{array}
\nonumber
\end{align}

\begin{align}
\begin{array}{c|l}
\cdot & \(21\)_{20} \\
\hline\hline
\(21\)_{20} & \(42\)_{86}+\(411\)_{46}+\(33\)_{31}+2\(321\)_{44}
      +\(3111\)_{6}+\(222\)_{10}+\(2211\)_{6}+ 	\\
      &+{\red2\(3\)_{19}+4\(21\)_{20}+2\(111\)_{4}+\(0\)_{1}} \\
\(111\)_{4} & \(321\)_{44}+\(3111\)_{6}+\(2211\)_{6}+
        {\red\(21\)_{20}+\(111\)_{4}} \\
\end{array}
\nonumber
\end{align}

\begin{align}
\begin{array}{c|l}
\cdot & \(111\)_{4} \\
\hline\hline
\(111\)_{4} & \(222\)_{10}+\(2211\)_{6} \\
\end{array}
\nonumber
\end{align}
\vfill\eject

\subsection{$\grpH_{21}(4)$ formal characters}

\paragraph{Dimensions of $\grpGL(4)\downarrow\grpH_{21}(4)$ irreps:}
\begin{align}
\begin{array}{|l|l|}
\hline
\{\lambda\}_{\textrm{dim}} & \(\lambda/M_{21}\)_{\textrm{dim}} \\ \hline
\{0\}_1 	&  \(0\)_1 \\
\{1\}_{4} 	&  \(1\)_4 \\
\{11\}_6 	&  \(11\)_6 \\
\{111\}_{4} 	&  \(111\)_4 \\
\{1^4\}_{1} 	&  \(1^4\)_{1} \\
\{2\}_{10} 	&  \(2\)_{10} \\
\{21\}_{20} 	&  \(21\)_{19}+\(0\)_{1} \\
\{211\}_{15} 	&  \(211\)_{11}+\(1\)_{4} \\
\{2111\}_{4} 	&  \(2111\)_{-2}+\(11\)_{6} \\
\{21^4\}_{0} 	&  \(21^4\)_{-4}+\(111\)_{4} \\
\{21^5\}_{0} 	&  \(21^5\)_{-1}+\(1^4\)_{1} \\
\{22\}_{20} 	&  \(22\)_{16}+\(1\)_{4} \\
\{221\}_{20} 	&  \(221\)_{4}+\(2\)_{10}+\(11\)_{6} \\
\{2211\}_{6} 	&  \(2211\)_{-17}+\(21\)_{19}+\(111\)_{4} \\
\{22111\}_{0} 	&  \(22111\)_{-12}+\(211\)_{11}+\(1^4\)_{1} \\
\{221^4\}_{0} 	&  \(221^4\)_{2}+\(2111\)_{-2} \\
\{2^21^5\}_{0} 	&  \(2^21^5\)_{4}+\(21^4\)_{-4} \\
\{2^21^6\}_{0} 	&  \(2^21^6\)_{1}+\(21^5\)_{-1} \\
\{3\}_{20} 	&  \(3\)_{20} \\
\{31\}_{45} 	&  \(31\)_{41}+\(1\)_{4} \\
\{311\}_{36} 	&  \(311\)_{20}+\(2\)_{10}+\(11\)_{6}  \\
\{31^3\}_{10} 	&  \(31^3\)_{-14}+\(21\)_{19}+\(111\)_{4}+\(0\)_{1} \\
\{31^4\}_{0} 	&  \(31^4\)_{-16}+\(211\)_{11}+\(1^4\)_{1}+\(1\)_{4} \\
\{31^5\}_{0} 	&  \(31^5\)_{-4}+\(21^3\)_{-2}+\(11\)_{6} \\
\{32\}_{60} 	&  \(32\)_{44}+\(2\)_{10}+\(11\)_{6} \\
\{321\}_{64} 	&  \(321\)_{1}+\(3\)_{20}+2\(21\)_{19}+\(111\)_{4}+\(0\)_{1} \\
\{3211\}_{20} 	& 
\(3211\)_{-68}+\(31\)_{41}+\(22\)_{16}+2\(211\)_{11}+\(1^4\)_{1}+2\(1\)_{4} \\
\{321^3\}_{0} 	&  
\(321^3\)_{-42}+\(311\)_{20}+\(221\)_{4}+2\(21^3\)_{-2}+\(2\)_{10}+2\(11\)_{6}\\
\{321^4\}_{0} 	& 
\(321^4\)_{12}+\(31^3\)_{-14}+\(2211\)_{-17}+2\(21^4\)_{-4}+\(21\)_{19}+2\(1^3\)_{4}\\
\{321^5\}_{0}  	&
\(321^5\)_{17}+\(31^4\)_{-16}+\(221^3\)_{-12}+2\(21^5\)_{-1}+\(211\)_{11}+2\(1^4\)_{1}
\\
\{321^6\}_{0}	&
\(321^6\)_{4}+\(31^5\)_{-4}+\(2^21^4\)_{2}+\(21^3\)_{-2} \\
\{33\}_{50} 	&  
\(33\)_{31}+\(21\)_{19} \\ \hline
\end{array}
\end{align}

\paragraph{Some examples of product formulae for $\grpH_{21}(4)$ characters:}

\begin{align}
\begin{array}{c|l}
\cdot & \(1\)_{4} \\
\hline\hline
\(1\)_{4}  & \(2\)_{10}+\(11\)_{6} \\
\(2\)_{10} & \(3\)_{20}+\(21\)_{19}+{\red\(0\)_{1}} \\ 
\(11\)_{6} & \(21\)_{19}+\(111\)_{4}+{\red\(0\)_{1}} \\
\(3\)_{20} & \(4\)_{35}+\(31\)_{41}+{\red\(1\)_{4}} \\
\(21\)_{19}& \(31\)_{41}+\(22\)_{16}+\(211\)_{11}+{\red2\(1\)_{4}} \\
\(111\)_{4}& \(211\)_{11}+\(1^4\)_{1}+{\red\(1)_{4}} \\ 
\end{array}
\nonumber
\end{align}

\begin{align}
\begin{array}{c|l}
\cdot & \(2\)_{10} \\
\hline\hline
\(2\)_{10}  & \(4\)_{35}+\(31\)_{41}+\(22\)_{16}+{\red2\(1\)_{4}} \\ 
\(11\)_{6}  & \(31\)_{41}+\(211\)_{11}+{\red2\(1\)_{4}} \\
\(3\)_{20}  & \(5\)_{56}+\(41\)_{74}+\(32\)_{44}+{\red2\(2\)_{10}+\(11\)_{6}} \\
\(21\)_{19} & \(41\)_{74}+\(32\)_{44}+\(311\)_{20}+\(221\)_{4}
       +{\red3\(2\)_{10}+3\(11\)_{6}} \\
\(111\)_{4} & \(311\)_{20}+{\green\(2111\)_{-2}}
       +{\red\(2\)_{10}+2\(11\)_{6}} \\ 
\end{array}
\nonumber
\end{align}

\begin{align}
\begin{array}{c|l}
\cdot & \(11\)_{6} \\
\hline\hline
\(11\)_{6} & \(22\)_{16}+\(211\)_{11}+\(1^4\)_{1}+{\red2\(1\)_{4}} \\
\(3\)_{20} & \(41\)_{74}+\(311\)_{20}+{\red2\(2\)_{10}+\(11\)_{6}} \\
\(21\)_{19}& \(32\)_{44}+\(311\)_{20}+\(221\)_{4}+{\green\(2111\)_{-2}}
      +{\red3\(2\)_{10}+3\(11\)_{6}} \\
\(111\)_{4}& \(221\)_{4}+{\green\(2111\)_{-2}}
       +{\red\(2\)_{10}+2\(11\)_{6}} \\ 
\end{array}
\nonumber
\end{align}

\begin{align}
\begin{array}{c|l}
\cdot & \(3\)_{20} \\
\hline\hline
\(3\)_{20}  & \(6\)_{84}+\(51\)_{120}+\(42\)_{86}+\(33\)_{31}
       +{\red2\(3\)_{20}+2\(21\)_{19}+\(0\)_{1}} \\ 
\(21\)_{19} & \(51\)_{120}+\(42\)_{86}+\(411\)_{31}+\(321\)_{1}
       +{\red3\(3\)_{20}+4\(21\)_{19}+\(111\)_{4}+2\(0\)_{1}} \\
\(111\)_{4} & \(411\)_{31}+{\green\(3111\)_{-14}}
        +{\red\(3\)_{20}+2\(21\)_{19}+\(111\)_{4}+\(0\)_{1}} \\ 
\end{array}
\nonumber
\end{align}

\begin{align}
\begin{array}{c|l}
\cdot & \(21\)_{19} \\
\hline\hline
\(21\)_{19} & \(42\)_{86}+\(411\)_{31}+\(33\)_{31}+2\(321\)_{1}
       +{\green\(3111\)_{-14}+\(222\)_{-10}+\(2211\)_{-17}}+ 	\\
      &+{\red4\(3\)_{20}+8\(21\)_{19}+4\(111\)_{4}+4\(0\)_{1}} \\
\(111\)_{4} & \(321\)_{1}+{\green\(3111\)_{-14}}+{\red\(3\)_{20}}
       +{\green\(2211\)_{-17}+\(21^4\)_{-4}}
       +{\red4\(21\)_{19}+3\(1^3\)_{4}+2\(0\)_{0}} \\
\end{array}
\nonumber
\end{align}

\begin{align}
\begin{array}{c|l}
\cdot & \(111\)_{4} \\
\hline\hline
\(111\)_{4} & {\green\(222\)_{-10}+\(2211\)_{-17} +\(21^4\)_{-4}}
              +{\red2\(21\)_{19}+2\(111\)_{4}+\(0\)_{1}} \\
\end{array}
\nonumber
\end{align}

\subsection{$\grpH_{1^3}(4)$ formal characters}
\label{subsec:grpH_1^3}
\paragraph{Dimensions of $\grpGL(4)\downarrow\grpH_{1^3}(4)$ irreps:}
\begin{align}
\begin{array}{|l|l|}
\hline
\{\lambda\}_{\textrm{dim}} & \(\lambda/M_{21}\)_{\textrm{dim}} \\ \hline
\{0\}_1 	&  \(0\)_1 \\
\{1\}_{4} 	&  \(1\)_4 \\
\{11\}_6 	&  \(11\)_6 \\
\{111\}_{4} 	&  \(111\)_3+\(0\)_{1} \\
\{1^4\}_{1} 	&  \(1^4\)_{-3}+\(1\)_{4} \\
\{1^5\}_{0} 	&  \(1^5\)_{-6}+\(11\)_{6} \\
\{1^6\}_{0} 	&  \(1^6\)_{-3}+\(1^3\)_{3} \\
\{1^7\}_{0} 	&  \(1^7\)_{3}+\(1^4\)_{-3} \\
\{2\}_{10} 	&  \(2\)_{10} \\
\{21\}_{20} 	&  \(21\)_{20} \\
\{211\}_{15} 	&  \(211\)_{11}+\(1\)_{4} \\
\{2111\}_{4} 	&  \(2111\)_{-12}+\(2\)_{10}+\(11\)_{6} \\
\{21^4\}_{0} 	&  \(21^4\)_{-24}+\(21\)_{20}+\(1^3\)_{3}+\(0\)_{1} \\
\{21^5\}_{0} 	&  \(21^5\)_{-12}+\(211\)_{11}+\(1^4\)_{-3}+\(1\)_{4} \\
\{21^6\}_{0} 	& 
\(21^6\)_{12}+\(21^3\)_{-12}+\(1^5\)_{-6}+\(11\)_{6} \\
\{22\}_{20} 	&  \(22\)_{20} \\
\{221\}_{20} 	&  \(221\)_{14}+\(11\)_{6} \\
\{2211\}_{6} 	&  \(2211\)_{-17}+\(21\)_{20}+\(1^3\)_{3} \\
\{221^3\}_{0} 	&  
\(221^3\)_{-32}+\(22\)_{20}+\(211\)_{11}+\(1^4\)_{-3}+\(1\)_{4}\\
\{221^4\}_{0} 	& 
\(221^4\)_{-12}+\(221\)_{14}+\(21^3\)_{-12}+\(2\)_{10}+\(1^5\)_{-6}+\(11\)_{6} \\
\{221^5\}_{0} 	& 
\(221^5\)_{21}+\(2211\)_{-17}+\(21^4\)_{-24}+\(21\)_{20}+\(1^6\)_{-3}+\(1^3\)_{3} \\
\{221^6\}_{0} 	& 
\(221^6\)_{33}+\(221^3\)_{-32}+\(21^5\)_{-12}+\(211\)_{11}+\(1^7\)_{3}+\(1^4\)_{-3} \\
\{2^3\}_{10} 	&  \(2^3\)_{6}+\(1^3\)_{3}+\(0\)_1 \\
\{2^31\}_{4} 	&  \(2^31\)_{-8}+\(211\)_{11}+\(1^4\)_{-3}+\(1\)_4 \\
\{2^31^2\}_{0} 	& 
\(2^31^2\)_{-8}+\(221\)_{14}+\(21^3\)_{-12}+\(1^5\)_{-6}+2\(11\)_{6} \\
\{2^31^3\}_{0} 	& 
\(2^31^3\)_{11}+\(2^3\)_{6}+\(2211\)_{-17}+\(21^4\)_{-24}+\(21\)_{20}+\(1^6\)_{-3}+2\(1^3\)_{3}+\(0\)_{1} \\
\{2^4\}_{1} 	&  \(2^4\)_{3}+\(21^3\)_{-12}+\(2\)_{10} \\ 
\hline
\end{array}
\end{align}

\paragraph{Some examples of product formulae for $\grpH_{1^3}(4)$ characters:}

\begin{align}
\begin{array}{c|l}
\cdot & \(1\)_{4} \\
\hline\hline
\(1\)_{4}  & \(2\)_{10}+\(11\)_{6} \\
\(2\)_{10} & \(3\)_{20}+\(21\)_{20} \\ 
\(11\)_{6} & \(21\)_{20}+\(111\)_{3}+\(1\)_{1} \\
\(3\)_{20} & \(4\)_{35}+\(31\)_{45} \\
\(21\)_{20}& \(31\)_{45}+\(22\)_{20}+\(211\)_{11}+{\red\(1\)_{4}} \\
\(111\)_{3}& \(21^2\)_{11}+{\green\(1^4\)_{-3}}+{\red\(1\)_{4}} \\ 
\end{array}
\nonumber
\end{align}

\begin{align}
\begin{array}{c|l}
\cdot & \(2\)_{10} \\
\hline\hline
\(2\)_{10}  & \(4\)_{35}+\(31\)_{45}+\(22\)_{30} \\ 
\(11\)_{6}  & \(31\)_{45}+\(211\)_{11}+{\red\(1\)_{4}} \\
\(3\)_{20}  & \(5\)_{56}+\(41\)_{84}+\(32\)_{60} \\
\(21\)_{20} & \(41\)_{84}+\(32\)_{60}+\(311\)_{26}+\(221\)_{14}
       +\(2\)_{10}+\(11\)_{6} \\
\(111\)_{3} & \(311\)_{26}+{\green\(2111\)_{-12}}
       +{\red\(2\)_{10}+\(11\)_{6}} \\ 
\end{array}
\nonumber
\end{align}

\begin{align}
\begin{array}{c|l}
\cdot & \(11\)_{6} \\
\hline\hline
\(11\)_{6} & \(22\)_{20}+\(211\)_{11}+{\green\(1^4\)_{-3}}+{\red2\(1\)_{4}} \\
\(3\)_{20} & \(41\)_{84}+\(311\)_{26}+{\red\(2\)_{10}} \\
\(21\)_{20}& \(32\)_{60}+\(311\)_{26}+\(221\)_{14}+{\green\(2111\)_{-12}}
      +{\red2\(2\)_{10}+2\(11\)_{6}} \\
\(111\)_{3}& \(221\)_{14}+{\green\(2111\)_{-12}+\(1^5\)_{-5}}
      +{\red\(2\)_{10}+2\(11\)_{6}} \\ 
\end{array}
\nonumber
\end{align}

\begin{align}
\begin{array}{c|l}
\cdot & \(3\)_{20} \\
\hline\hline
\(3\)_{20}  & \(6\)_{84}+\(51\)_{140}+\(42\)_{126}+\(33\)_{50} \\ 
\(21\)_{20} & \(51\)_{140}+\(42\)_{126}+\(411\)_{50}+\(321\)_{44}
       +{\red\(3\)_{20}+\(21\)_{20}} \\
\(111\)_{3} & \(411\)_{50}+{\green\(3111\)_{-30}}
        +{\red\(3\)_{20}+\(21\)_{20}} \\ 
\end{array}
\nonumber
\end{align}

\begin{align}
\begin{array}{c|l}
\cdot & \(21\)_{20} \\
\hline\hline
\(21\)_{20} & \(42\)_{126}+\(411\)_{50}+\(33\)_{50}+2\(321\)_{44}
      +{\green\(3111\)_{-30}}+{\red2\(3\)_{20}}+\(222\)_{6}+ \\
     &+{\green\(2211\)_{-17}}+{\red4\(21\)_{20}+2\(111\)_{3}+\(0\)_{1}} \\
\(111\)_{3} & \(321\)_{44}+{\green\(3111\)_{-30}}+{\red\(3\)_{20}}
      +{\green\(2211\)_{-17}+\(2211\)_{-20}} \\
     &+{\red3\(21\)_{20}+2\(111\)_{3}+\(0\)_{1}} \\
\end{array}
\nonumber
\end{align}

\begin{align}
\begin{array}{c|l}
\cdot & \(111\)_{3} \\
\hline\hline
\(111\)_{3} & \(222\)_{6}+{\green\(2211\)_{-17}+\(21^4\)_{-20}}
       +{\red2\(21\)_{20}}+{\green\(1^6\)_{-3}}+{\red2\(1^3\)_{3}+\(0\)_{1}} \\
\end{array}
\nonumber
\end{align}

\end{appendix}
\small{

}
\end{document}